\documentclass[twocolumn]{aastex631}

\usepackage{color}
\usepackage{amsmath}

\def\ergs{\textit{$\rm erg\ s^{-1}$}}
\def\loglbol{\textit{${\rm log}\,L_{\rm bol}$}}
\def\lbol{\textit{$L_{\rm bol}$}}
\def\edd{\textit{$\lambda_{\rm Edd}$}}
\def\mo{\textit{$m_{\rm o}$}}
\def\lambdao{\textit{$\lambda_{\rm o}$}}

\def\smo{\textit{$\sigma_m$}}
\def\sso{\textit{$\sigma_s$}}

\def\so{\textit{$s_{\rm o}$}}
\def\logedd{\textit{${\rm log}\,\lambda_{\rm Edd}$}}
\def\nobs{\textit{$N_{\rm obs}$}}
\def\pac{\textit{$p_{\rm ac}$}}

\def\mbh{\textit{$\mathcal{M}_{\rm BH}$}}
\def\m{\textit{$\mathcal{M}_{\star}$}}

\def\mbulge{\textit{$\mathcal{M}_{\rm bulge}$}}
\def\logm{\textit{${\rm log}\,\m$}}
\def\logmbh{\textit{${\rm log}\,\mbh$}}

\def\deltalogmbh{\textit{$\Delta\,\logmbh$}}

\def\rel{\textit{$\mbh-\m$}}
\def\bulgerel{\textit{$\mbh-\mbulge$}}

\def\hb{\textit{${\rm H}\beta$}}
\def\ha{\textit{${\rm H}\alpha$}}
\def\cii{\textsc{[C$\,$ii]}}

\def\lim{\textit{$l_{\rm lim}$}}

\def\msun{\textit{$M_\odot$}}
\def\ergs{\textit{$\rm erg\ s^{-1}$}}
\def\kms{\textit{$\rm km\ s^{-1}$}}

\begin{document}

\title{Tip of the iceberg: overmassive black holes at $4<z<7$ found by JWST are not inconsistent with the local \rel\ relation}


\author{Junyao Li}
\affiliation{Department of Astronomy, University of Illinois at Urbana-Champaign, Urbana, IL 61801, USA}
\correspondingauthor{Junyao Li}
\email{junyaoli@illinois.edu}

\author{John D. Silverman}
\affiliation{Kavli Institute for the Physics and Mathematics of the Universe, The University of Tokyo, Kashiwa, Japan 277-8583 (Kavli IPMU, WPI)}
\affiliation{Department of Astronomy, School of Science, The University of Tokyo, 7-3-1 Hongo, Bunkyo, Tokyo 113-0033, Japan}
\affiliation{Center for Data-Driven Discovery, Kavli IPMU (WPI), UTIAS, The University of Tokyo, Kashiwa, Chiba 277-8583, Japan}
\affiliation{Center for Astrophysical Sciences, Department of Physics \& Astronomy, Johns Hopkins University, Baltimore, MD 21218, USA}

\author{Yue Shen}
\affiliation{Department of Astronomy, University of Illinois at Urbana-Champaign, Urbana, IL 61801, USA}
\affiliation{National Center for Supercomputing Applications, University of Illinois at Urbana-Champaign, Urbana, IL 61801, USA}

\author{Marta Volonteri}
\affiliation{Institut d'Astrophysique de Paris, Sorbonne Universit\'e, CNRS, UMR 7095, 98 bis bd Arago, 75014 Paris, France}

\author{Knud Jahnke}
\affiliation{Max-Planck-Institut für Astronomie, Königstuhl 17, D-69117 Heidelberg, Germany}

\author{Ming-Yang Zhuang}
\affiliation{Department of Astronomy, University of Illinois at Urbana-Champaign, Urbana, IL 61801, USA}

\author{Matthew T. Scoggins}
\affiliation{Department of Astronomy, Columbia University, New York, NY, 10027}

\author{Xuheng Ding}
\affiliation{School of Physics and Technology, Wuhan University, Wuhan 430072, China}

\author{Yuichi Harikane}
\affiliation{Institute for Cosmic Ray Research, The University of Tokyo, 5-1-5 Kashiwanoha, Kashiwa, Chiba 277-8582, Japan}

 \author{Masafusa Onoue}
 \affiliation{Kavli Institute for the Physics and Mathematics of the Universe, The University of Tokyo, Kashiwa, Japan 277-8583 (Kavli IPMU, WPI)}
 \affiliation{Center for Data-Driven Discovery, Kavli IPMU (WPI), UTIAS, The University of Tokyo, Kashiwa, Chiba 277-8583, Japan}
 \affiliation{Kavli Institute for Astronomy and Astrophysics, Peking University, Beijing 100871, P.R.China}

\author{Takumi S. Tanaka}
\affiliation{Department of Astronomy, School of Science, The University of Tokyo, 7-3-1 Hongo, Bunkyo, Tokyo 113-0033, Japan}
\affiliation{Kavli Institute for the Physics and Mathematics of the Universe, The University of Tokyo, Kashiwa, Japan 277-8583 (Kavli IPMU, WPI)}
\affiliation{Center for Data-Driven Discovery, Kavli IPMU (WPI), UTIAS, The University of Tokyo, Kashiwa, Chiba 277-8583, Japan}

\begin{abstract}
\noindent JWST is revealing a new remarkable population of high-redshift ($z\gtrsim4$), low-luminosity Active Galactic Nuclei (AGNs) in deep surveys and detecting the host galaxy stellar light in the most luminous and massive quasars at $z\sim 6$ for the first time. Latest results claim supermassive black holes (SMBHs) in these systems to be significantly more massive than expected from the local BH mass -- stellar mass  (\rel) relation and that this is not due to sample selection effects. Through detailed statistical modeling, we demonstrate that the coupled effects of selection biases (i.e., finite detection limit and requirements on detecting broad lines) and measurement uncertainties in \mbh\ and \m\ can in fact largely account for the reported offset and flattening in the observed \rel\ relation toward the upper envelope of the local relation, even for those at \mbh\ $< 10^8\,M_{\odot}$. We further investigate the possible evolution of the \rel\ relation at $z\gtrsim 4$ with careful treatment of observational biases and consideration of the degeneracy between intrinsic evolution and dispersion in this relation. The bias-corrected intrinsic \rel\ relation in the low-mass regime suggests that there might be a large population of low-mass BHs ($\logmbh\lesssim5$), possibly originating from lighter seeds, remaining undetected or unidentified even in the deepest JWST surveys. These results have important consequences for JWST studies of BH seeding and the coevolution between SMBHs and their host galaxies at the earliest cosmic times.
\end{abstract}


\section{Introduction}

The James Webb Space Telescope (JWST) is revolutionizing the study of high redshift AGNs and the early coevolution of SMBHs with their host galaxies. The unprecedented sensitivity and resolution of NIRCam imaging have enabled the first detection of stellar light from AGNs during the reionization epoch ($z\sim6$) and estimates of their stellar masses. With NIRSpec, BH mass estimates, based on the broad \ha\ and \hb\ lines, at such high redshifts are routinely measured with ease. Consequently, several recent studies have investigated the \rel\ relation at early cosmic times \citep[e.g.,][]{Kocevski2023, Maiolino2023, Harikane2023, Ding2023, Stone2023, Yue2023}. This relation, with its slope, normalization, and intrinsic scatter being sensitive to the merger history of galaxies and the impact of AGN feedback, has emerged as one of the most important scaling relations in unraveling the nature of SMBH--galaxy coevolution \citep[e.g.,][]{Peng2007, Volonteri2009, Hirschmann2010, Jahnke2011, Kormendy2013, Li2021mass, Ding2022simu, Habouzit2022, LiJ2023, Zhuang2023}. Additionally, the \rel\ relation, close to the seeding epochs ($z\sim10-30$), may shed light on the formation mechanisms of SMBHs \citep[e.g.,][]{Volonteri2009, Volonteri2010, Trinca2022, Scoggins2023}.

In contrast to expectations based on the tight \bulgerel\ relation in the local universe \citep[e.g.,][]{Kormendy2013}, AGNs currently found at $4<z<7$ tend to significantly lie above this local relation and exhibit larger scatter in the \rel\ plane. The detection of ``overmassive'' BHs (i.e., elevated above the mean local \rel\ relation), at least to some extent, is anticipated given a well-understood selection bias \citep{Lauer2007}, since only luminous and massive BHs with emission sufficiently dominating their host galaxy can be observed in a flux-limited survey \cite[e.g.,][]{Treu2007, Schulze2011, Volonteri2011, Shen2015, Sun2015, Li2021mass, Li2022, Zhang2023, Volonteri2023, Tanaka2024}. This selection bias is suggested to potentially explain the $\gtrsim 1$~dex offset (relative to the local relation) found for $z\sim6$ quasars by \cite{Schulze2014} and \cite{Li2022}, although their analyses are based on \cii-traced dynamical mass measured by ALMA. 
Contrary to this, several recent JWST studies have argued that selection bias is either negligible in deep JWST surveys or insufficient to explain the large observed offset at high redshift \citep[e.g.,][]{Pacucci2023, Maiolino2023, Stone2023}. 

In particular, \cite{Pacucci2023} estimate that the observed \rel\ relation at $z\sim5$ for low-luminosity AGNs discovered in the deep JADES \citep{Eisenstein2023} and CEERS \citep{Finkelstein2023} surveys is a factor of $\sim10-100$ above that of local AGNs and the bias due to selection is 0.2~dex upwards at most. Consequently, they conclude that the overall population of high redshift BHs is intrinsically overmassive compared to their low-$z$ counterparts.
Reaching a similar conclusion, \cite{Stone2023} compare the distribution of $z\sim6$ quasars to 30 PG quasars ($z<0.5$) and find that these two samples occupy distinct regions in the \rel\ plane. This is treated as evidence that the elevated BH mass seen in high-redshift quasars cannot be attributed to selection effects.

If selection effects were minimal, particularly at the lower luminosities probed by deep JWST fields, the prevalence of overmassive observed AGNs during the reionization epoch suggests that the majority of the high-redshift BH population is overmassive. This finding could support the theory that SMBHs mainly originate from heavy seeds via direct collapse BHs  ($\sim10^{5}\,\msun$; e.g., \citealt{Begelman2006, Agarwal2013, Visbal2018, Pacucci2023, Bogdan2024}), with SMBHs initially much more massive than the budding host galaxy \citep{Agarwal2013}.

However, the impact of selection effects, measurement uncertainties and appropriate choice of the local relation are not yet well established to exclude the existence of undermassive BHs grown from light seeds ($\sim100\,\msun$, formed from the remnants of massive Population III stars; \citealt{Fryer2001}) or medium-weight seeds ($\sim1000\,\msun$, formed from runaway stellar collisions in dense star clusters; \citealt{Madau2001}), since they may not have grown enough to be detectable \citep{Natarajan2017, Volonteri2023}. Specifically, \cite{Pacucci2023} adopted a constant detection limit for \mbh, derived from the \ha\ line sensitivity, in their bias analysis. In principle, the detection limit on \mbh\ in a flux-limited survey should depend on the Eddington ratio (\edd) distribution function (ERDF) of the AGN population. A lower {\it{intrinsic}} \edd\ would result in only massive BHs being observable. 
Of similar concern, the PG quasars, used as a low-redshift benchmark in \cite{Stone2023}, have lower luminosities and a different selection function compared to high redshift quasars, due to being drawn from different mass, accretion rate, and duty cycle distribution functions at low and high redshifts \citep{Schulze2011}.

In this paper, we conduct a detailed statistical modeling of selection biases for a sample of high redshift AGNs observed by JWST, with a focus on recently discovered low-luminosity AGNs, to understand the impact of selection effects on revealing the intrinsic connection between SMBH growth and galaxy evolution (i.e., the \rel\ relation) in the early universe ($z\sim 4-7$). Furthermore, we address the use of different local relations as the baseline, particularly between active and non-active samples. 
Throughout this paper we adopt a flat $\Lambda$CDM cosmology with $\Omega_{\Lambda}=0.7$ and $H_{0}=70\,{\rm km\,s^{-1}\,Mpc^{-1}}$. Magnitudes are given in the AB system. Stellar masses are matched to a \cite{Chabrier2003} initial mass function (IMF).

\section{Sample}
\label{sec:sample}

In this work, we use 32 AGNs at $z>4$, including luminous quasars, from the following samples with JWST observations.

\begin{itemize}

    \item Twelve low-luminosity AGNs as reported by \cite{Maiolino2023}\footnote{We do not include the  secondary BHs of the three dual AGN candidates identified through two broad \ha\ components, as their stellar mass is unknown and their nature is yet to be confirmed.}, and ten low-luminosity AGNs from \cite{Harikane2023} at $4<z<7$. These low luminosity AGNs, with bolometric luminosities and BH masses in the range of $44.1\lesssim\loglbol/\ergs \lesssim46.0$ and $6.5\lesssim \logmbh/\msun \lesssim 8.3$, were recently discovered in deep JWST surveys (hereafter the JDEEP sample), including the ERO \citep{Pontoppidan2022}, CEERS \citep{Finkelstein2023}, GLASS \citep{Treu2022}, and JADES \citep{Eisenstein2023} programs. 
    They are selected based on the detection of the broad component of the \ha\ or \hb\ lines by NIRSpec, requiring, for instance, a fullwidth at half maximum (FWHM) $\gtrsim 1000\ \kms$, with additional criteria for line detection significance and narrow line width to avoid line broadening due to outflows.
    
    \item Two moderate-luminosity quasars with $\loglbol/\ergs \sim46.2$ and $\logmbh/\msun \sim8.3-9.1$ from the SHELLQs survey \citep{Matsuoka2016} as reported in \cite{Ding2023}. These two quasars are part of the JWST Cycle 1 program (Observation ID 1967; PI: M. Onoue) which targets 12 of the moderate-luminosity quasars with $45.8\lesssim\loglbol/\ergs \lesssim46.4$ and $7.5\lesssim\logmbh/\msun\lesssim9.1$ at $6.0<z<6.4$ discovered by the Hyper Suprime-Cam Subaru Strategic Program \citep[HSC-SSP;][]{Aihara2022}.

    \item Ten luminous quasars in \cite{Stone2023} and \cite{Yue2023} at $5<z<7$, which probe the most luminous and massive BHs in the early universe ($\loglbol/\ergs \gtrsim47.2$ and $\logmbh/\msun\gtrsim8.3$). 
    These quasars are specifically selected to cover a range of  luminosity, star formation rate, and the properties of the $\rm Ly\alpha$ forest and Gunn–Peterson trough to study the early growth of SMBHs,  intergalactic medium, and cosmic reionization. Note that one of the quasars in \cite{Stone2023}, HSC J2239+0207, is selected from the SHELLQs survey, with its luminosity ($\loglbol/\ergs\sim46.4$) being an order of magnitude smaller than the rest of the sample. Therefore, we attribute it to the SHELLQs sample in the following analysis. 
\end{itemize}

BH masses are derived using the single-epoch virial estimator \citep{Vestergaard2006, Shen2013} from the broad \ha\ or \hb\ line. For consistency, we recalibrated the \ha-based BH mass in \cite{Harikane2023} using the updated \cite{Reines2015} recipe, as adopted in \cite{Maiolino2023}. 
The bolometric luminosities of low-luminosity AGNs are derived from the broad \ha\ luminosity ($L_{\ha}$) using the $L_{\ha}-L_{5100}$ relation in \cite{Greene2005} assuming $\lbol=10.3\times L_{5100}$ \citep{Richards2006}. The contamination of host galaxy to $L_{5100}$ is negligible for luminous quasars, thus their \lbol\ are directly estimated from the continuum luminosity at 5100\,\AA.
Stellar masses are determined through various methods: NIRSpec (plus NIRCam photometry) spectral decomposition in \cite{Maiolino2023}, 2D AGN+host image decomposition using NIRCam data in \cite{Harikane2023}, \cite{Ding2023}, and \cite{Yue2023}, as well as 1D surface brightness profile scaling as described in \cite{Stone2023}. 

We assume that these BH and stellar mass estimators are unbiased in the mean with a statistical uncertainty of $\sim0.4$~dex. However, in practice, the specific BH mass recipes adopted by these studies may be systematically biased (high) by $\sim0.1-0.2$ dex \citep{Shen2013, Shen2023}, especially for highly-accreting systems \citep{Du2016}.  The stellar mass estimates are  most likely biased due to systematic effects in spectra- or imaging-based AGN-host decomposition, especially for AGN-dominated systems at faint magnitudes \cite[][]{Li2021size, LiJ2023, Zhuang2023psf}, imperfect alignment of objects in the micro-shutter that causes substantial slit loss \citep{Tacchella2023}, older stars being overwhelmed by ongoing star formation \citep{Topping2022, Narayanan2023, Gimenez2023, Arteaga2024}, and uncertainties in the star formation history and IMF at early cosmic times \citep{Topping2022, Steinhardt2022, Whitler2023, Arteaga2024}. The existence of such biases can be tentatively seen in the distinct \m\ distributions for the \cite{Harikane2023} and  \cite{Maiolino2023} samples, with a $p$-value = 0.007 for a KS test and a mean (median) difference of 0.43 (0.75) dex, despite having similar distributions in \mbh\ and \lbol.

It is also noteworthy that, although the stellar masses for many undetected host galaxies are reported as upper limits in \cite{Harikane2023}, \cite{Stone2023}, and \cite{Yue2023}, these systems might actually be ``massive'' (compared to the upper limits) but compact galaxies that are indistinguishable from a point source \citep{Li2021size, Zhuang2023psf}. This effect is particularly pronounced for quasar-dominated systems and barely-resolved low-mass ($\m<10^{9}\ \msun$) galaxies at $z\gtrsim5$, where the typical galaxy size for the latter ($\lesssim0.5$~kpc, $\lesssim0.\arcsec09$) is smaller than the PSF in most long wavelength filters \citep{Maiolino2023, Baggen2023}. In principle, dedicated image and spectral energy distribution simulations are required to understand how reliable the host properties can be recovered in such systems \citep[][]{Li2021size, Zhuang2023psf}.

\begin{figure}
\centering
\includegraphics[width=\linewidth]{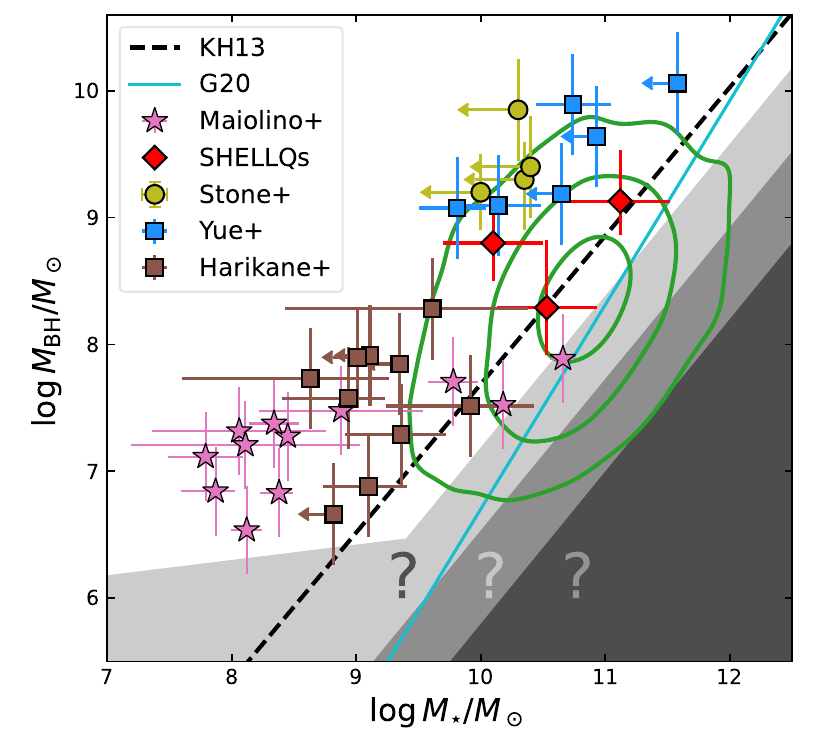}
\caption{The \rel\ relation of high-redshift AGNs used in this study \citep{Harikane2023, Maiolino2023, Stone2023, Yue2023,Ding2023}. For comparison, we display the SDSS quasar sample ($\loglbol/\ergs\lesssim46$) at $0.2<z<1$ in \cite{Li2021size, Li2021mass} as green contours ($1\,\sigma-3\,\sigma$ level), along with the local \bulgerel\ relation in \cite{Kormendy2013} and the local \rel\ relation in \cite{Greene2020}. The gray shaded regions display a schematic view of the inaccessible region of high-redshift observations, where the BHs are either undetectable or under-represented (sparsely sampled) due to the limited sensitivity  and survey volume, with the exact limits depending on specific survey designs.}
\label{fig:sample}
\end{figure}

With these caveats in mind, we present the \rel\ relation of our compiled high-redshift AGN sample in Figure \ref{fig:sample}.
The distinct distribution of various samples demonstrates how surveys with varying area and depth complicate our understanding of this relation. The rare and high-luminosity quasars in massive host galaxies come from individual pointings pre-selected from $\sim$all-sky but shallow surveys \citep{Fan2023}, which are only capable of discovering the most massive systems. The wide and deep HSC-SSP survey has extended the detection limit to somewhat lower BH masses \citep{Matsuoka2016, Onoue2019}, while the very low-luminosity ones are detected in deep but pencil-beam JWST fields that mainly probe common low-mass galaxies. This explains why the low-luminosity samples do not recover any low-mass BHs residing in massive galaxies, since the latter population are sparse on the sky due to the bottom-heavy galaxy stellar mass function (SMF; e.g., \citealt{Weaver2023}).

\section{\rel\ relation at high-$z$}
\label{sec:mass_relation}

To elucidate the connection between the early growth of SMBHs and their host galaxies, a common approach is to compare their location in the \rel\ plane to that observed in the local universe. In this study, we adopt the following local relations (Figure \ref{fig:sample}): (1) The \bulgerel\ relation with an intrinsic scatter of $\sim0.3$~dex from \cite{Kormendy2013} (KH13). This relation -- established with massive, (classical) bulge-dominated quiescent galaxies with dynamical BH mass measurements -- has been the standard for  exploring SMBH--galaxy coevolution. Note that KH13 increased the mass-to-light ratio by 0.125 dex to match the dynamical zero point when measuring the bulge mass. We utilize their unadjusted relation to maintain consistency with our AGN sample. 
(2) The \rel\ relation for ``all'' the local massive BHs ($2 < \logmbh < 10$) from \cite{Greene2020} (G20). This relation, measured based on the total stellar mass rather than the bulge mass and exhibiting a much lower normalization and larger scatter ($\sim0.81$~dex) than the KH13 relation, is established through a combined fit of local early- and late-type galaxies with (mostly) dynamical BH mass measurements.

While the observed high redshift AGNs do exhibit an elevated $\mbh/\m$ in comparison to the mean local ratios in Figure \ref{fig:sample}, the dependence of \mbh\ on \lbol\ ensures that the \cite{Lauer2007} bias, induced by flux limits, and intrinsic scatter of the mass relation must be present. There will always be some regions in the high-$z$ \rel\ plane that are either not accessible to any observations due to the finite detection limit, or that are much sparsely populated in any dataset due to the coupled effects of the limited survey area, the scatter of the mass relation, and the shape of the SMF. A schematic view of this inaccessible region is displayed in Figure \ref{fig:sample} as gray shaded regions. The exact limits depend on the specific sample in question and can vary strongly with population and redshift.

In Figure~\ref{fig:sample}, we also display the \rel~relation for $\sim4000$ SDSS quasars at $0.2<z<1.0$ and above a redshift-dependent stellar mass from \cite{Li2021size, Li2021mass} as green contours. In contrast to the small sample comparison in \cite{Stone2023}, we find that the distribution of high-redshift quasars is covered by the upper envelope of the SDSS quasar sample. Therefore, ``overmassive'' BHs exist in both high- and low-redshift universe. This is true even considering that the low-$z$ quasar sample in \cite{Li2021size} only probes $\loglbol/\ergs \lesssim 46$ and misses the most luminous and massive quasars that are sparse on the sky, due to their requirement of having HSC-SSP DR2 coverage ($\rm \sim300\,deg^2$) for measuring  stellar masses. 
The lack of $z\sim6$ quasars populating the lower envelope of the SDSS sample might be attributed to the shallower detection limit at high redshift, resulting in the undetection of the entire ``undermassive'' population.
In the following sections, we conduct a rigorous modeling of observational biases to substantiate this hypothesis.

\section{Method}
\label{sec:method}
We quantify the impact of selection effects and measurement uncertainties on the intrinsic mass relation by modeling the conditional probability of observing a \mbh\ for a given \m\ at a specified detection limit. We then infer the intrinsic mass relation from the observed biased sample through forward modeling as widely adopted in the literature \cite[e.g.,][]{Schulze2011, Shen2015, Sun2015, Li2021mass, Li2022, LiJ2023, Ding2023, Tanaka2024}. Here we outline the statistical recipes for measuring the bivariate distribution of the observed \mbh\ and \m. 

We define the true mass terms as $m\equiv\logmbh$ and $s\equiv\logm$, while the observed mass measurements with Gaussian uncertainties \smo\ and \sso\ included are defined as \mo\ and \so, respectively. The luminosity of the quasar $l\equiv \loglbol$ is related to $m$ through the Eddington ratio $\lambda\equiv \logedd$ as $l= \lambda + m + 38.1$. We ignore the uncertainty on $l$ which is on the order of $\sim0.04$~dex \citep{Maiolino2023, Harikane2023}, although we note that it only considers errors propagated from the emission line or continuum flux and does not take into account the uncertainty associated with the bolometric conversion factor. The observed Eddington ratio \lambdao\ can then be measured as $l - \mo - 38.1$. 

At a given true $m$, the line width ($w\equiv \rm { log}\,FWHM$, assuming negligible uncertainty) of the broad line (BL) follows a Gaussian distribution $g(w)$ with the mean determined by the virial estimator \citep{Vestergaard2006} as 
\begin{equation}
\begin{split}
w = 0.5\,m - 0.25\,l + 10.8.
\label{eq:fwhm}
\end{split}
\end{equation}
Its scatter $\sigma_w$ (0.14 dex), along with the scatter in $l$ at a fixed $m$ (i.e., the scatter of the ERDF, which is about 0.4~dex; \citealt{Wu2022}), gives rise to the uncertainty in the virial BH mass estimate (0.35 dex) as
\begin{equation}
\smo = \sqrt{(0.5 \sigma_l)^2 + (2 \sigma_w)^2}.
\label{eq:sigmaw}
\end{equation}

We assume that $m$ is correlated with $s$ through a linear relation $m = \alpha s + \beta$ with a slope $\alpha$, normalization $\beta$, and a mass-independent Gaussian intrinsic scatter $\sigma$, i.e., 
\begin{equation}
g(m|s) = \frac{1}{\sqrt{2\pi}\sigma}\, {\rm exp}\Bigg(-\frac{(m - (\alpha s + \beta))^2}{2\sigma^2}\Bigg).
\label{eq:mbh_m_prob}
\end{equation}
Note that for mass relations with a steep slope, a small normalization, or/and a large scatter, the BH mass below the average for small galaxies could be negative. In such cases, the Gaussian distribution of \mbh\ at a given \m\ can be considered as setting all values where $\mbh < 100\ \msun$ to $\mbh = 100\ \msun$ (i.e., a truncated Gaussian distribution with a second peak at $\mbh=100$), which results in the same probability of observing overmassive BHs as the original Gaussian distribution. The physical picture under such a mass distribution could be the existence of a significant population of light seeds that fail to grow efficiently \citep[e.g.,][]{Smith2018}.

The probability of detecting an AGN at a redshift of $z$, with true BH mass $m$, stellar mass $s$, and luminosity $l$, can be expressed as 
\begin{equation}
\begin{aligned}[b]
p(m, s, l, z) &= \psi(s, z) \times \frac{dV}{dz}\\
& \times g(m|s, z) \times p_{\rm ac} (m, z) \times p(\lambda, z)\\
& \times \Omega(m, \lambda, w, z),
\end{aligned}
\label{eq:prob_detect}
\end{equation}
where $\psi(s, z)$ is the SMF, $p_{\rm ac}$ is the active 
unobscured BL AGN fraction, $p({\lambda}, z)$ is the ERDF, and $\Omega(m, \lambda, w, z)$ is the selection function of observations. 

We attribute the selection function primarily to the  AGN luminosity limit ($l_{\rm lim}$) achievable in a specific survey design, as well as the requirement of having line widths $\gtrsim1000\ \kms$ \citep[e.g.,][]{Harikane2023}, which inevitably excludes low-mass BHs whose BLs are intrinsically narrower. 
The luminosity limit is determined as the maximum value between $L_{\rm phot, lim}$ and $L_{\rm spec, lim}$. Here, $L_{\rm phot, lim}$ is the minimum \lbol\ above which an AGN can be photometrically selected as AGN candidates for scheduling spectroscopic follow-up observations (e.g., for the SHELLQs and Quasar samples). The $L_{\rm spec, lim}$ is determined by the emission line sensitivity of spectroscopic observations and the requirement for robustly identifying the BLs, which can be converted into a detection limit on \lbol\ via its tight correlation with line luminosity \citep{Greene2005}. 

For simplicity, we consider the lowest luminosity observed for each sample ($\loglbol/\ergs=44.1$, 46.2, and 47.2, for the JDEEP, SHELLQs, and Quasar samples, respectively), as representative of the luminosity limit. A more detailed treatment of the selection function would take into account the host galaxy contamination and the complex photometric selection criteria and target priority given in the shutter allocation (which may be biased towards or against AGNs; \citealt{Maiolino2023}), as well as additional requirements on line detections and dedicated criteria to achieve certain science goals. However, such an analysis is beyond the scope of the current work. 
Our approach of adopting an {\it{effective}} luminosity limit  based on that of spectroscopically-confirmed AGNs effectively reflects the current capability to detect and identify BL AGNs with specific survey designs and color selection method. 

In our model, BHs can be detected and identified as BL AGNs if their luminosity $l$ exceed the detection limit $l_{\rm lim}$ and the line width exceed $1000$\,\kms. Therefore, the selection function on $m$ can be expressed as 
\begin{equation}
\begin{aligned}[b]
    \Omega(m) =  &\ \pac(m)\, g_w(w>3|l>l_{\rm lim})\\
    & \int_{l_{\rm lim}}^{l_{\rm max}} p_{\lambda} (l-m-38.1)\, dl.
\end{aligned}
\label{eq:selection}
\end{equation}
Given the lack of observational constraints, by default we assume that \pac\ does not vary with \mbh, and thus it can be removed from the above Equation. The potential impact of a mass-dependent \pac\ is demonstrated in Appendix~\ref{sec:appendix}.
In general, a positive correlation between \pac\ and \mbh\ at high redshift (e.g, due to increased fueling efficiency) tends to bias the observed mass relation of AGNs towards massive BHs, even without any selection on luminosity. This is in contrast to the situation in the local universe, where less massive BHs are preferentially observed in AGN samples due to the reduced accretion activity for the most massive BHs \citep{Schulze2011}.

The bivariate distribution function of the observed \mo\ and \so, which incorporates measurement uncertainties and the selection function, is formulated as 
\begin{equation}
    p(\mo, \so) = \int{g(\mo|m)\, g(\so|s)\, p(m,s,z)\, dm\, ds}.
\label{eq:p_mo_so}
\end{equation}
The observed \rel\ relation (i.e., the expectation value of \mo\ at a given \so) can then be derived as 
\begin{equation}
\mo(\so) = \frac{\int{\mo\,p(\mo, \so)\,d\mo}}{\int{p(\mo, \so)\,d\mo}}.
\label{eq:mo_so}
\end{equation}

Measuring the bivariate \rel\ distribution and the expected ``observed'' mass relation given the selection function and mass uncertainties relies upon prior knowledge of the underlying {\it{intrinsic}} distribution functions (see Eq. \ref{eq:prob_detect}). In this work, we adopt the latest SMF from \cite{Carrera2023} which utilized deep JWST data to constrain the slope at the low-mass end. As for the ERDF, we use the fiducial result presented in \cite{Wu2022} (i.e., integrated over $8.5 < \logmbh/\msun < 10.5$, which can be approximated by a log-normal distribution with a mean of $\sim-1.0$ and a scatter of $\sim0.4$), who derived the intrinsic ERDF for quasars at $z\sim6$ with selection effects corrected through a Bayesian framework. 

The BL AGN fraction $p_{\rm ac}$ remains uncertain at $z\sim5$. 
The fraction of BL AGNs with $L_{\ha}>2.3\times10^{42}\ \ergs$ ($\loglbol/\ergs>44.7$) to UV-selected galaxies in the \ha\ flux-limited EIGER and FRESCO NIRCam/WFSS surveys is $\sim1\%$ at $-19<M_{\rm UV}<-22$ ($8\lesssim\logm/\msun\lesssim10$; \citealt{Song2016}). This fraction increases to $\sim5\%$ in the deeper CEERS + GLASS surveys \citep{Harikane2023}, and further rises to $\sim10\%$ in the deepest JADES survey \citep{Maiolino2023}, although it may be affected by the complicated survey selection function. 
Conservatively, the fraction of all the unobscured BL AGNs considered by our model (i.e., including those below the flux limit) must exceed $\sim1\%$. On the other hand, the AGN duty cycle (equivalent to the active fraction if the BH occupation fraction is 100\%) at $z\sim6$, deduced from solving the continuity equation, ranges from $\sim15\%$ at $\logmbh/\msun\sim6.5$ and increases to $60\%$ for $\logmbh/\msun\sim8.3$ \citep{Aversa2015}. 
This duty cycle can be converted into the unobscured BL AGN fraction by assuming an unobscured/obscured ratio. Despite existing uncertainties and sample incompleteness, the ratio of unobscured to obscured AGN number is $\sim1:2$ in the JADES survey \citep{Maiolino2023, Scholtz2023}, broadly consistent with the constraint from X-ray observations \citep{Vito2018}. 
Based on these findings, we consider $\pac \sim 1\%-15\%$ (assumed to be mass-independent) in this study. The impact of the assumed \pac\ range is discussed in Section \ref{subsec:intrinsic}.

\begin{figure*}
\centering
\includegraphics[width=0.4\linewidth]{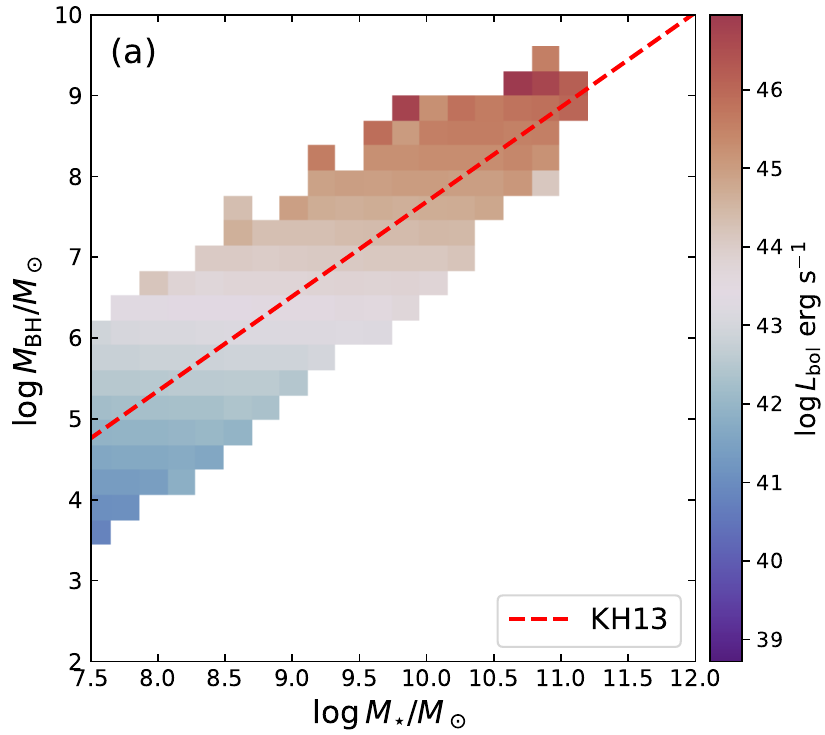}
\includegraphics[width=0.4\linewidth]{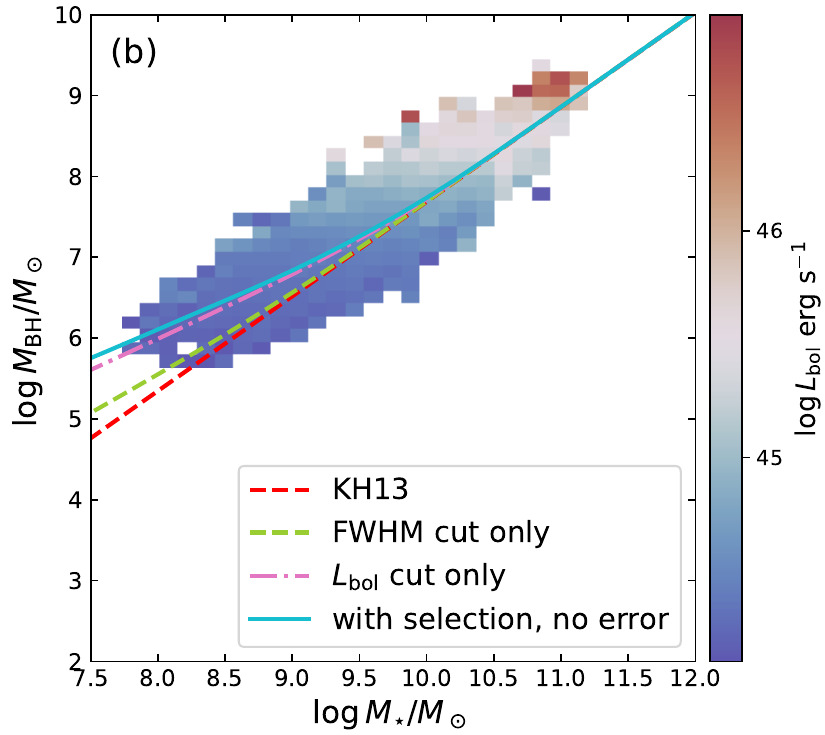}
\includegraphics[width=0.4\linewidth]{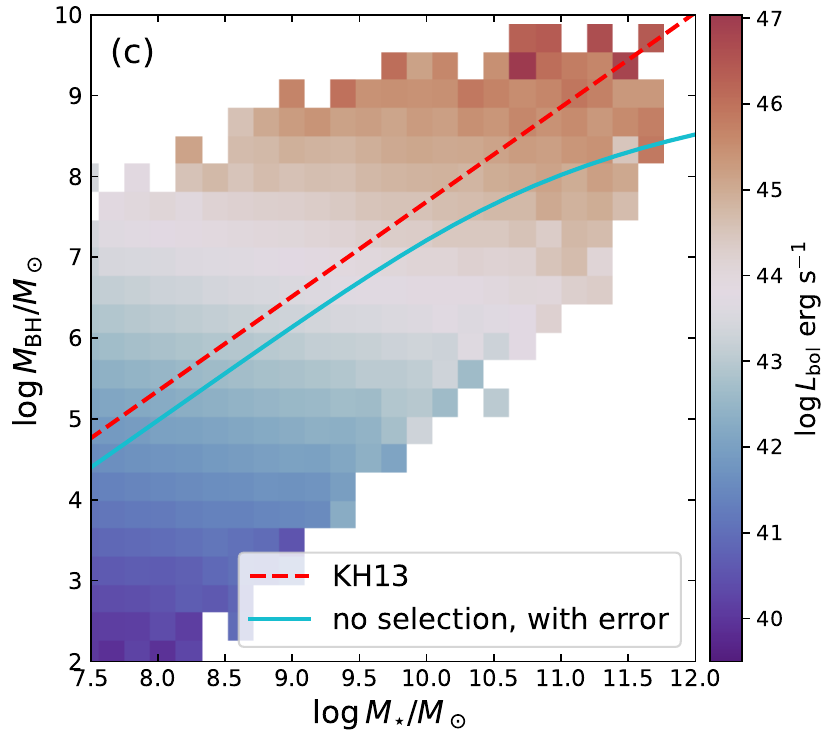}
\includegraphics[width=0.4\linewidth]{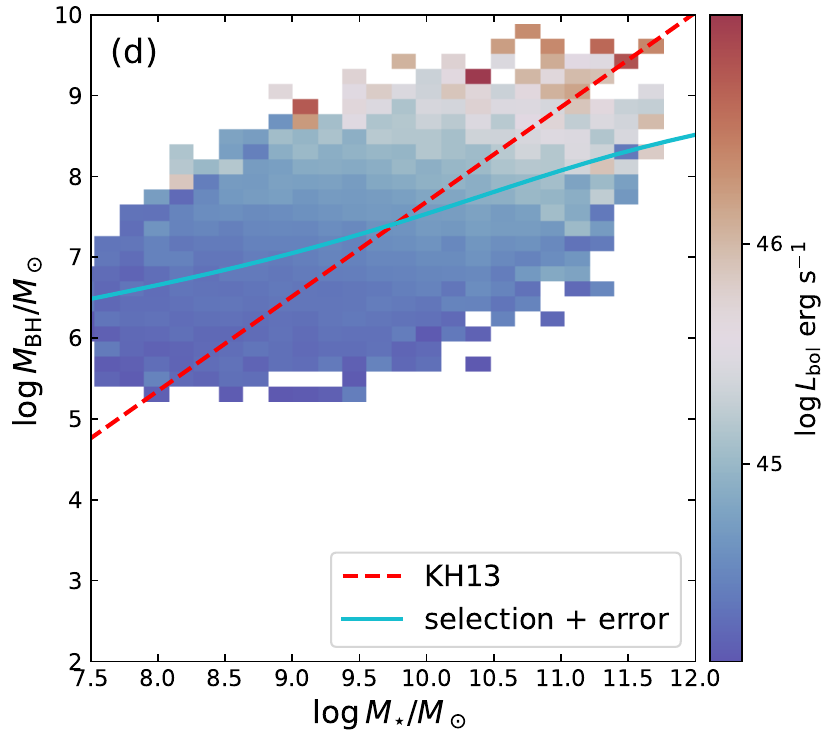}
\caption{Impact of selection effects and measurement uncertainties on the \rel\ relation. The square data points, color-coded by the average \lbol\ in each mass bin, represent mock AGNs simulated using Equation \ref{eq:p_mo_so} based on the local KH13 relation (red dashed line) with an intrinsic scatter of 0.3 dex. The cyan solid curve represents the mean \rel\ relation averaged over the square data points. {\bf{Panel (a)}}: The original KH13 relation. 
{\bf{Panel (b)}}: no mass uncertainties added but with the selection function ($\loglbol>44.1\ \ergs$ and $\rm FWHM > 1000\ \kms$) applied to the input KH13 relation. The effect of applying the $\lbol$ and the FWHM cut separately is displayed for comparison.
{\bf{Panel~(c)}}: mass uncertainties added on both \mbh\ (0.35 dex) and \m\ (0.45 dex), but without the selection function applied.
{\bf{Panel~(d)}}:  mass uncertainties added and selection function applied. }
\label{fig:sample_simu}
\end{figure*}

\section{Results}
\label{sec:bias}

\subsection{Impact of observational biases}
\label{subsec:impact}

We first demonstrate how the observational biases (i.e., selection effect induced by flux limits and measurement uncertainties) modify the intrinsic \rel\ relation using a Monte Carlo simulation that mimics the {\it{effective}} JDEEP observations. We assume that SMBHs at $z=5.2$ (the mean redshift of the JDEEP sample) follow the local KH13 relation with an intrinsic scatter of 0.3 dex. By randomly sampling Equation~\ref{eq:p_mo_so}, we create mock AGNs with known intrinsic $\mbh$, $\m$, $\lbol$ and line width. The distribution of the resulting mock sample in the \rel\ plane, color-coded by the average \lbol\ in each mass bin, is displayed in Figure \ref{fig:sample_simu}a. It can be seen that \mbh\ strongly depends on \lbol, which is the fundamental cause of the selection bias. 

In Figure \ref{fig:sample_simu}b we display the mass relation after applying the selection criteria of $\loglbol/\ergs>44.1$ and $\rm FWHM>1000\ \kms$ to the mock AGNs in Figure \ref{fig:sample_simu}a. It exhibits a significant positive offset at the low-\m\ end, due to the \cite{Lauer2007} bias of missing low-mass BHs with lower luminosities. The offset is mainly caused by the luminosity cut, while the FWHM cut further excludes a small portion of bright AGNs whose line width is narrower than 1000\,\kms\ (Figure \ref{fig:sample_simu}b).

In Figure \ref{fig:sample_simu}c we incorporate the average mass uncertainties for the JDEEP sample ($\smo = 0.35$ and $\sso = 0.45$) to the mock AGNs in Figure \ref{fig:sample_simu}a to illustrate the impact of measurement uncertainties. We caution that $\sso = 0.45$ is likely an underestimate, given that the reported stellar mass uncertainty for some sources can be as small as 0.03 dex, possibly due to the small flux error adopted in SED fitting. The actual uncertainties, considering e.g., template and SFH mismatch, are expected to be much larger. Notably, the distribution with added mass uncertainties is significantly broader than the original KH13 relation. Moreover, its mean relation appears to shift downwards compared to the KH13 relation. 
This shift is attributed to the substantial stellar mass uncertainties and the bottom-heavy SMF, leading to the \cite{Eddington1913} bias. To elaborate, consider a stellar mass bin at $\logm/\msun\sim8.5$. Given the abundance of small BHs hosted by low-mass galaxies with $\logm/\msun < 8.5$, the added uncertainties on the stellar mass lead to numerous low-mass BHs scattering rightwards unevenly into the $\logm/\msun \sim 8.5$ bin. This effect surpasses the leftward scattering of the rarer high-mass BHs with $\logm/\msun > 8.5$ into the same stellar mass bin. Consequently, the mean {\it{observed}} BH mass at $\logm/\msun \sim 8.5$ is biased low.

Finally, we apply the $\loglbol/\ergs>44.1$ and $\rm FWHM>1000\ \kms$ criteria to mock AGNs in Figure \ref{fig:sample_simu}c.
This gives rise to the {\it{observed}} \rel\ relation which incorporates both realistic mass uncertainties and selection function as real observations. As shown in Figure~\ref{fig:sample_simu}d, it appears noticeably flatter than the KH13 relation, exhibiting a significant positive offset at the low-\m\ end and a negative offset at the high-\m\ end, due to the combined effect of the Lauer bias and the Eddington bias. 
The comparison between Figures \ref{fig:sample_simu}b and  \ref{fig:sample_simu}d demonstrates that substantial mass uncertainties further amplify the observed offset in \mbh/\m\ at the low-\m\ end (up to $\sim1.5$ dex), rather than simply broadening the distribution. This is because, when adding stellar mass uncertainties to mock AGNs in Figure \ref{fig:sample_simu}b, where low-mass BHs have been excluded by the selection criteria, the average observed \mbh\ in the $\logm/\msun \sim 8.5$ bin is actually dominated by massive BHs at $\logm/\msun>8.5$ scattering leftwards into it. This situation differs from the one discussed earlier for Figure \ref{fig:sample_simu}c.

These analyses demonstrate that it is essential to consider the overall transformation in the observed \rel\ distribution (i.e., deviation, broadening, and flattening relative to the intrinsic mass relation) caused by selection bias and measurement uncertainties, even in deep JWST surveys like JADES and CEERS.

\subsection{JWST-observed overmassive AGNs}

\label{subsec:observed_offset}

\begin{figure*}
\centering
\includegraphics[width=\linewidth]{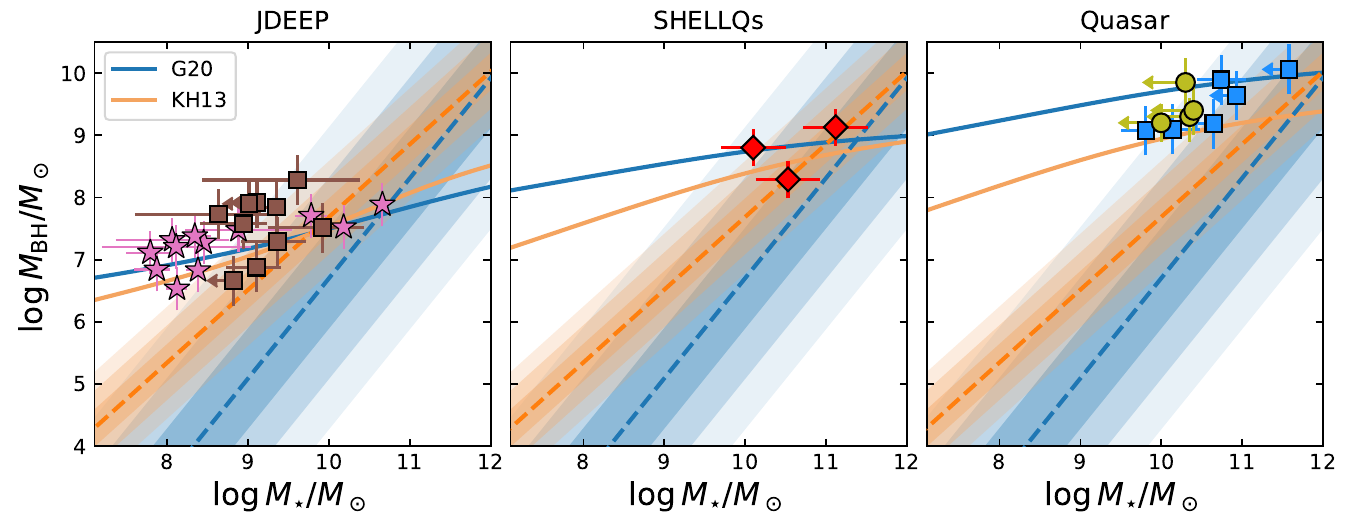}
\caption{The predicted ``observed'' \rel\ relations (solid curves) of detectable AGNs at $z\sim5$ in the three surveys, assuming that the underlying BH population (i.e., including those below the detection limits) intrinsically follow the local KH13 and G20 relations (dashed curves with the same colors). The intrinsic scatter at $1\,\sigma$, $2\,\sigma$, and $3\,\sigma$ levels of each local relation are shown as shaded regions. Only BHs located in the upper envelope of the assumed intrinsic mass relations can be detected as luminous AGNs. Their positions are further biased upwards by the substantial mass uncertainties (Figure \ref{fig:sample_simu}), resulting in a significantly elevated \mbh/\m\ compared to the intrinsic ratio. 
}
\label{fig:bias}
\end{figure*}

Based on the results shown in Figure \ref{fig:sample_simu}, we apply the selection function and the average measurement uncertainties of each observed sample to the local KH13 and G20 relations. This approach enables us to predict the \rel\ relation that we would expect to observe at the effective detection limit of each survey from Equation \ref{eq:mo_so}, assuming that SMBHs at $z\sim5$ intrinsically follow the local relations. Intriguingly, as shown in Figure \ref{fig:bias}, in all scenarios, the predicted \rel\ relations above the luminosity limit deviate significantly from the assumed local relations and generally follow the observed data points within the observed stellar mass ranges. This holds true even for the most luminous quasars exhibiting a $\gtrsim 1$~dex offset, confirming earlier studies based on ALMA \citep{Schulze2014, Li2022}.
It is also interesting to note that although the G20 relation has a lower average \mbh/\m\ compared to the KH13 relation, its predicted observed \mbh/\m\ could be even higher due to its larger scatter, resulting in an increased number of observable overmassive BHs. Note that the scatter of the G20 relation might be mass-dependent and decrease at the massive end, but this is not considered in their fitting.

These results imply that, even if the underlying $z\sim5$ BH population follows the local relation on average, we would only observe its upper envelope, namely overmassive AGNs, due to the finite detection limit of JWST or/and the specific sample selection criteria. Our analysis emphasizes the importance of properly considering the dispersion of the mass relation when investigating its evolution, rather than simply comparing averages across cosmic times.
Figure \ref{fig:bias} also exemplifies that one should avoid plotting a dataset against a given intrinsic correlation without transforming the latter to match the dataset's selection and measurement uncertainties. Failing to do so can easily lead to misinterpretations. 

\begin{figure*}
\centering
\includegraphics[width=0.49\linewidth]{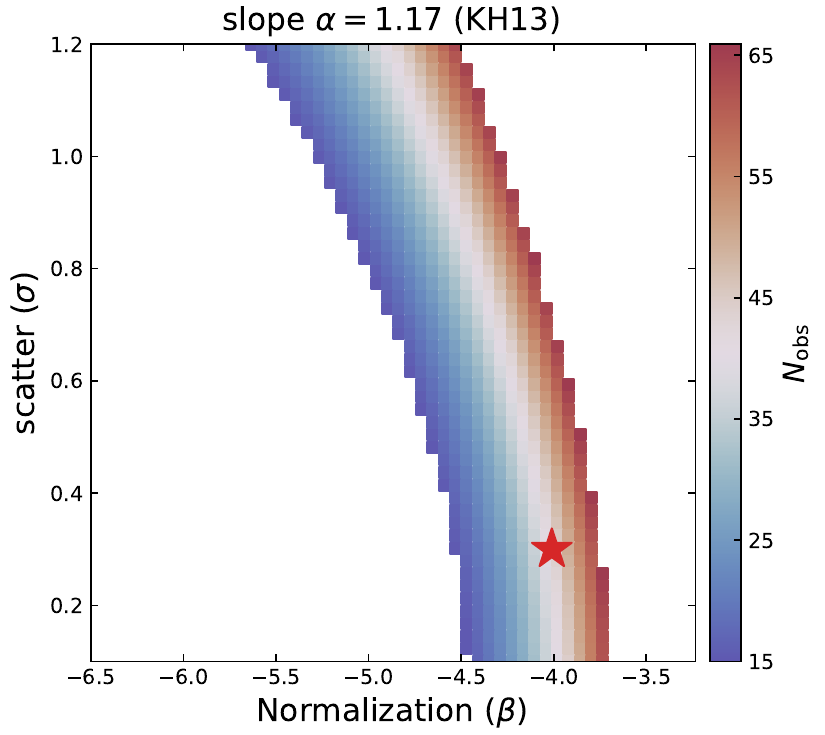}
\includegraphics[width=0.49\linewidth]{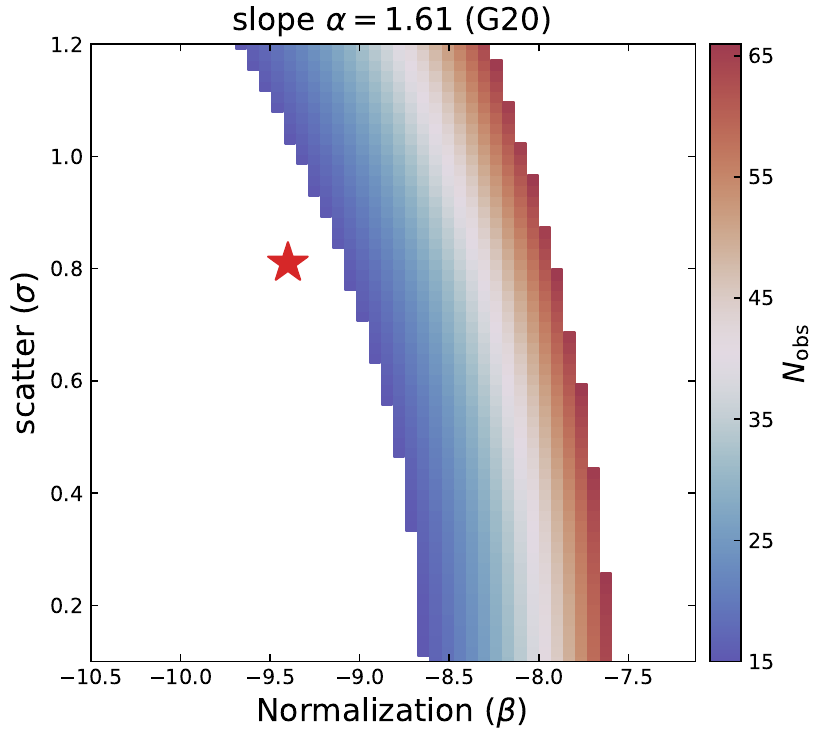}
\caption{Model-predicted permitted region ($15<\nobs<65$ for AGNs with $\loglbol>44.1$, $7.8<\logm<10.6$, and $4<z<7$) in the $\beta-\sigma$ plane for $\alpha=1.17$ (KH13) and $\alpha=1.61$ (G20), respectively. The values of $\beta$ and $\sigma$ for the KH13 and G20 relations are marked by the red stars. Each bin is color-coded by its average \nobs. The forbidden region shown in white indicates parameter spaces in $[\alpha, \beta, \sigma]$ that result in either too few ($<15$, left side) or too many ($>65$, right side) observable AGNs compared to observations ($\nobs=22$), assuming $\pac=10\%$. }
\label{fig:observable}
\end{figure*}

It is noteworthy that the predicted \rel\ relation only represents an idealized scenario for an infinitely large survey, as the survey volume, BL AGN fraction, and the normalization of the SMF in Equation \ref{eq:prob_detect} cancel out in Equation \ref{eq:mo_so}. In practice, only BHs in common galaxies (stellar mass range determined by the survey area and depth) can be observed. As a result, specific parameter spaces in [$\alpha$, $\beta$, $\sigma$, \pac, \m, \mbh] are unobservable in small-area surveys, meaning that the expected number of AGNs above the detection limit is less than one in that survey. {In other words, the observed number counts of AGNs in a given survey impose additional constraints on the underlying intrinsic \rel\ relation and other distribution functions. While fully utilizing this AGN number count constraint in modeling the underlying relations is beyond the scope of the current work, we use this argument as a sanity check on the validity of the proposition that high-redshift SMBHs intrinsically follow the KH13 or the G20 relations.

We first adopt the JDEEP survey as an example to illustrate this point. Specifically, assuming different forms of the intrinsic \rel\ relation and BL AGN fraction, we count the number (\nobs) of observable AGNs (i.e., $\loglbol/\ergs>44.1$ and $\rm FWHM>1000\ \kms$) in a survey area of $\sim100$~arcmin$^2$ (8 NIRSpec pointings for CEERS and 8 pointings for JADES; \citealt{Harikane2023, Maiolino2023, Eisenstein2023})\footnote{The effective area for each NIRSpec pointing used for shutter target placement in JADES is $\sim7$~arcmin$^2$. Taking into account pointing overlaps, we adopt a total area of $\sim50$~arcmin$^2$ for JADES and assume the same area for CEERS.} within the observed stellar mass range ($7.8<\logm/\msun<10.6$) at $4<z<7$. 
If the model produces a number of observable BL AGNs commensurate to observations (i.e., 22 sources in JDEEP), we consider it as a permitted model. Otherwise, it is deemed forbidden, meaning that specific combinations of $\alpha$, $\beta$, $\sigma$, and \pac\ either produces too many bright AGNs that are not observed, or fail to produce any observable ones. 
In practice, we adopt $15<\nobs<65$ to identify permitted models.
This somewhat arbitrary \nobs\ range is determined by first adjusting the actual observed AGN counts by a factor of 1.5 (i.e., $\nobs=15-33$) to accommodate the uncertainty in the exact galaxy number counts in our adopted SMF and in converting \mbh\ to \lbol\ via the assumed ERDF (a higher \edd\ would result in more observable AGNs). We further increase the upper limit by an additional factor of two (i.e., $\nobs=15-65$) to account for potential missing BL AGNs in observations that exceed the luminosity limit but remain unidentified due to various reasons (e.g., outshined by the host galaxy, restricted shutter arrangement). 

As shown in Figure \ref{fig:observable}, the KH13 relation falls well within the permitted region ($\nobs\sim40$) under $\pac=10\%$. The G20 relation exhibits a slight deviation ($\nobs\sim10$) if we adopt $\pac=10\%$, but a slightly higher (or broader) mass relation or \pac\ could bring it back into agreement. 
Similarly, we count the number of observable quasars with $\loglbol/\ergs>47.2$ and $9.8<\logm/\msun<11.5$ over half of the sky \citep{Fan2023} across $5<z<7$, assuming $\pac=10\%$. The predicted number based on the KH13 relation is $\nobs \sim 56$, in good agreement with the few tens of quasars of comparable luminosities discovered thus far \citep{Fan2023}. The G20 relation, with a mean similar to the KH13 relation at the massive end and a much larger scatter (0.81 dex), predicts $\nobs \sim 568$. This demonstrates that the intrinsic scatter for massive galaxies cannot be that large if the norm and slope are those of G20, unless a significant population of unobscured quasars were missed by the current color selection methods.

In general, both the KH13 and G20 relations can generate a sufficient number of observable overmassive AGNs under reasonable assumptions and reproduce the observed flat \rel\ relation given the sample selection function (Figure \ref{fig:bias}), even if their mean \mbh/\m\ are significantly smaller than the observed ratio. This analysis illustrates that the reported offset in the \rel\ plane can be replicated by simply incorporating observational biases (including measurement uncertainties) into the local relations (i.e., by sampling their $\gtrsim2\,\sigma$ upper envelope), even assuming that the mass estimates are unbiased on average.

\subsection{Intrinsic \rel\ relation for JWST-discovered low-luminosity AGNs}
\label{subsec:intrinsic}

\begin{figure*}
\centering
\includegraphics[width=0.48\linewidth]{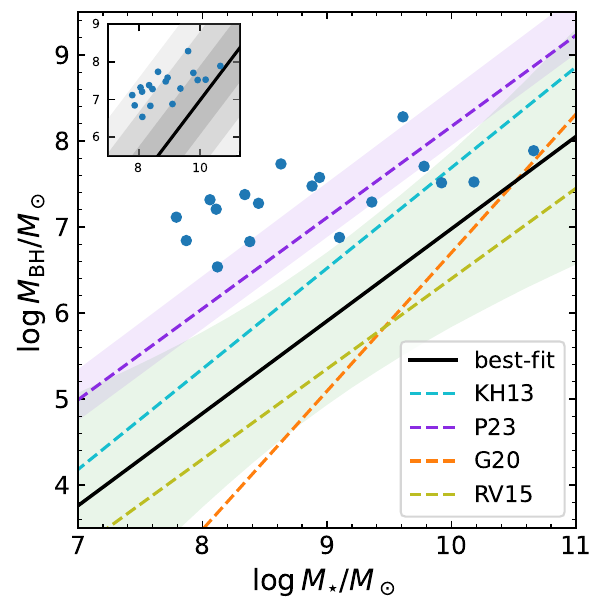}
\includegraphics[width=0.5\linewidth]{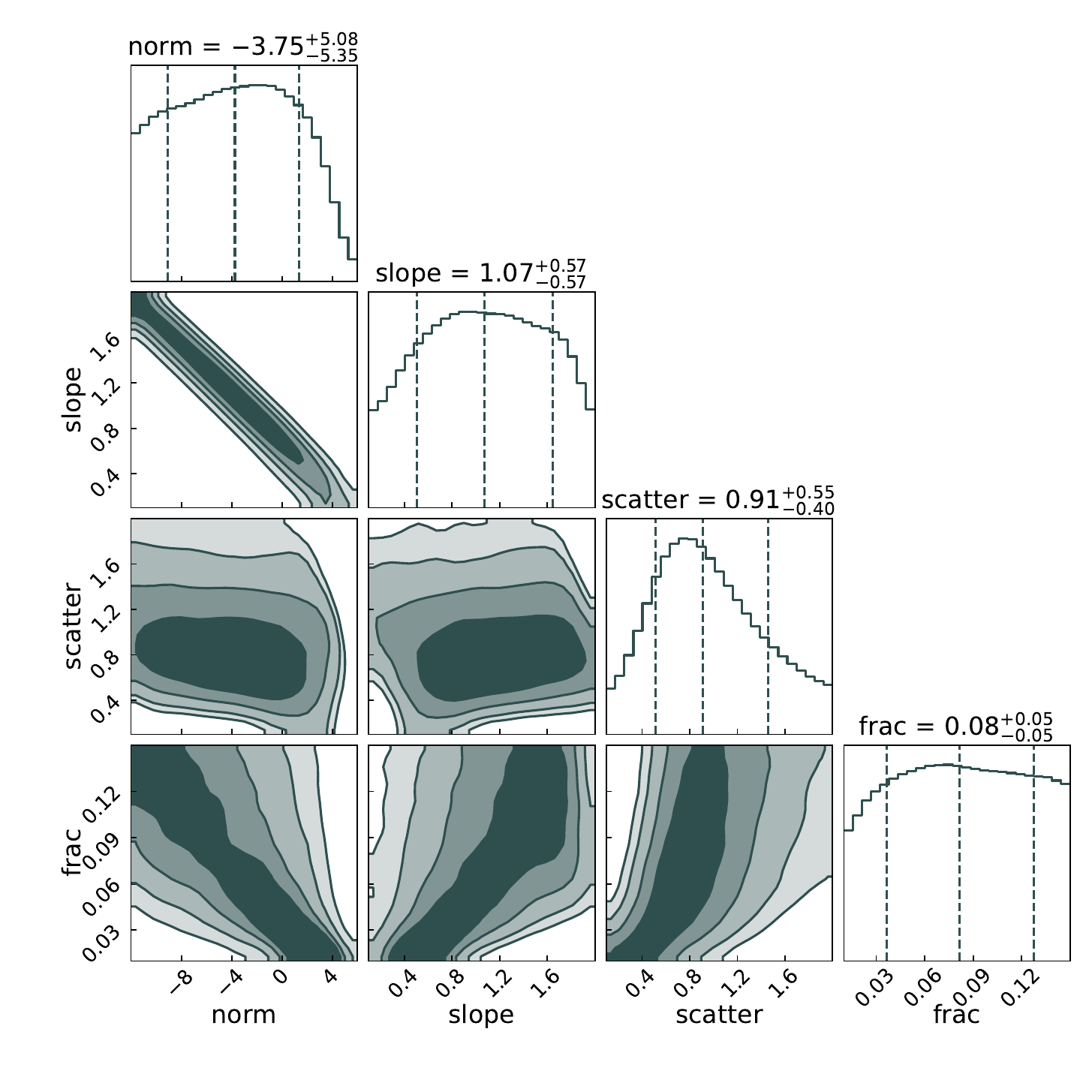}
\caption{ {\bf Left:} best-fit intrinsic \rel\ relation (black line) and its $1\,\sigma$ uncertainty (green shaded region) inferred from the JDEEP sample with solid host galaxy detections (blue points). The inset panel shows the expanded $1-3\,\sigma$ intrinsic scatter (gray shaded regions). The intrinsic \rel\ relation for the same sample derived by \cite{Pacucci2023} and its $1\,\sigma$ uncertainty are shown in magenta. Local relations from KH13, G20, and RV15 are also shown for comparison. {\bf Right:} posterior parameter distributions and best-fit values of the mass relation and BL AGN fraction for the observed AGN sample show on the left.} 
\label{fig:best_fit}
\end{figure*}

To differentiate whether the observed \rel\ relation is drawn from the KH13, G20, or other forms of intrinsic mass relations that may appear similarly flat after applying the selection function (Figure \ref{fig:bias}), we employ {\tt{emcee}} to sample the conditional probability distribution $p(\mo|\so)$ (i.e., the likelihood function) derived from Equation \ref{eq:p_mo_so} as functions of $\theta = [\alpha, \beta, \sigma, \pac]$. This enables us to infer the most likely intrinsic, bias-corrected \rel\ relation. That is, once convolved with measurement uncertainties and the sample selection function, this relation gives rise to the observed AGN sample. 

In this work, we focus on deriving the intrinsic \rel\ relation for the JDEEP sample and comparing our results with those in \cite{Pacucci2023}.  Similar analyses for a larger sample of 20 quasars within comparable mass and luminosity ranges as those of \cite{Stone2023} and \cite{Yue2023} have been conducted in \cite{Li2022} (albeit based on ALMA dynamical masses), who concluded that the biased-corrected \mbh/\m\ is consistent with a range of local values considering the uncertainties. A similar conclusion is reached in \cite{Ding2023} based on the pilot study of the first two SHELLQs quasars observed by JWST.

In the fitting, we only consider the 18 objects with robust host galaxy detections, as the stellar masses for ``undetected'' hosts are highly uncertain and not necessarily upper limits. In principle, the fitting methodology should be able to self-consistently reproduce the AGN luminosity function  given the ERDF, the best-fit \rel\ relation,  and AGN fraction. However, in practice this is infeasible due to the small sample size, non-uniform selection function, uncertain underlying distribution functions and their unknown mass dependency.
Therefore, we adopt a simplified approach by considering the integrated AGN number counts in the fitting as a constraint to eliminate forbidden parameters. Specifically, we define $P({\rm I}/\theta)=1$ if the model predicts $15<\nobs<65$, and $P({\rm I}/\theta)=0$ otherwise. We adopt flat priors for all parameters ($-12<\alpha<6$ , $0.1<\beta<2.0$, $0.1<\sigma<2.0$, and $1\%<\pac<15\%$). We then minimize the product of the likelihood function, prior, and $P({\rm I}/\theta)$ to compute the posterior parameter distributions.

\begin{table*}
\renewcommand{\arraystretch}{1.2}
\centering
\caption{Model fitting results in the form of $\logmbh = \alpha \times \logm + \beta$ with a Gaussian intrinsic scatter $\sigma$ and a BL AGN fraction \pac\ (Columns 3--6) derived under different integration ranges for \pac\ and \nobs\ (Columns 1--2). The fitting results under our default model assumptions are shown in bold. The average offset in BH mass at $7.8<\logm<10.6$ relative to the default result and that of \cite{Pacucci2023} are shown in Columns 7 and 8, respectively.}
\begin{tabular}{cccccccc}
\hline
\hline
\pac\ range & \nobs\ range & $\alpha$ & $\beta$ & $\sigma$ & $\pac$ & $\deltalogmbh$ (default) & $\deltalogmbh$ (P23)\\
(1) & (2) & (3) & (4) & (5) & (6) & (7) & (8)\\
\hline
$\mathbf{1\%-15\%}$ & $\mathbf{15-65}$ & $\mathbf{1.07_{-0.57}^{+0.57}}$ & $\mathbf{-3.75_{-5.35}^{+5.08}}$ & $\mathbf{0.91_{-0.40}^{+0.55}}$ & $\mathbf{0.08_{-0.05}^{+0.05}}$ & -- & $\mathbf{-1.20}$\\
$10\%-15\%$ & $15-65$ & $1.19_{-0.59}^{+0.50}$ & $-5.38_{-4.56}^{+5.13}$ & $1.04_{-0.39}^{+0.53}$ &  $0.12_{-0.02}^{+0.02}$ & $-0.48$ & --1.68\\
$1\%-15\%$ & $40-65$ & $0.97_{-0.51}^{+0.62}$ & $-2.51_{-5.68}^{+4.68}$ & $0.80_{-0.37}^{+0.52}$ & $0.09_{-0.05}^{+0.04}$ & $+0.28$ & --0.92\\
$10\%-15\%$ & $40-65$ & $1.19_{-0.6}^{+0.53}$ & $-4.87_{-4.79}^{+5.18}$ & $0.93_{-0.35}^{+0.49}$ & $0.12_{-0.02}^{+0.02}$ & $+0.04$ & --1.16\\

\hline
\end{tabular}
\label{table:params}
\end{table*}

Figure \ref{fig:best_fit} presents the best-fit intrinsic \rel\ relation and the corresponding posterior parameter distributions. The best-fit relation is parameterized as $\logmbh = 1.07_{-0.57}^{+0.57}\, \logm - 3.75_{-5.35}^{+5.08}$ with an intrinsic scatter of $0.91_{-0.40}^{+0.55}$ dex, corresponding to an observable BL AGN number of $31.0_{-12.0}^{+19.6}$. As expected, the slope and normalization of the intrinsic relation appear notably steeper and smaller than the observed one, after correcting for the selection bias of missing low-mass BHs. 
The observed AGNs with $\logm/\msun \gtrsim 9$ are positioned within the upper $1\,\sigma$ intrinsic scatter region (see the inset in Figure \ref{fig:best_fit}). While those with even higher BH masses are detectable, their number density is low, resulting in their underrepresentation in pencil-beam surveys. Conversely, the detected AGNs at $\logm/\msun<9$ (where the intrinsic \mbh/\m\ is lower) lie outside the $1\,\sigma$ region, as only the upper BH envelope is detectable. 
This result suggests that the overmassive AGNs detected by JWST are fully consistent with being drawn from an underlying BH population whose average BH mass is much smaller (by $\sim1.2$ dex) than what has been observed. 

The large scatter of $0.91_{-0.40}^{+0.55}$ dex in the low-mass regime, compared to that of the KH13 relation (with a probability of $p(\sigma>0.3) \sim 95.3\%$) and massive quasars at $z<2$ ($\sim0.3$ dex; \citealt{Li2021mass, Tanaka2024}), might suggest a merger-averaging origin for reducing the scatter and building the tight relationship between massive BHs and bulges \citep{Peng2007, Jahnke2011}.  This picture aligns with the recent discovery of a high dual AGN fraction \citep{Perna2023} and the prevalence of companion galaxies around BL AGNs \citep{Matthee2023} at high redshift. 

However, due to the limited sample size, substantial measurement uncertainties, and significant degeneracy among various parameters, the recovered intrinsic relation is subject to considerable uncertainties such that both the KH13 and G20 relations fall within its $1\,\sigma$ uncertainty region. Moreover, adjusting the integration range for \pac\ and \nobs\ also affects the fitting result, as certain forms of mass relations cannot reproduce the assumed AGN number counts. As summarized in Table~\ref{table:params}, restricting \pac\ to a lower range (e.g., $1\%-15\%$) or a higher range (e.g., $10\%-15\%$) leads to an upward or a downward shift, respectively, in the best-fit relation, while adjusting the \nobs\ range in a similar manner ($\nobs = 15-65$ vs. $40-65$) leads to the opposite effect. On the other hand, if a significant population of BL AGNs at $\loglbol>44.1$ is missed, the lower limit of the integral for both \pac\ and \nobs\ naturally increases (e.g., $\pac=10\%-15\%$ and $\nobs = 40-65$). This may mitigate its impact on the best-fit mass relation as the two parameters act in  opposite directions.

\section{Discussion}
\label{sec:discussion}

\subsection{Comparison with Previous Work}
\label{subsec:compare}

\begin{figure*}
\centering
\includegraphics[width=0.8\linewidth]{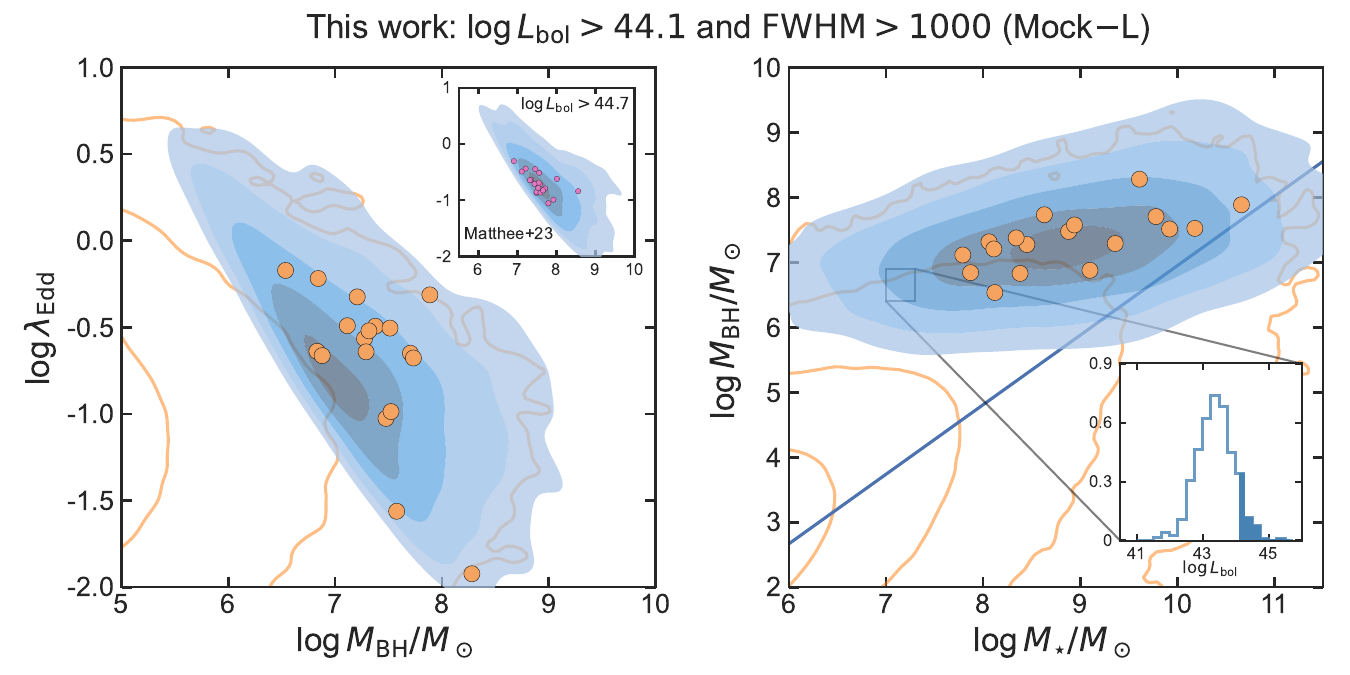}
\includegraphics[width=0.8\linewidth]{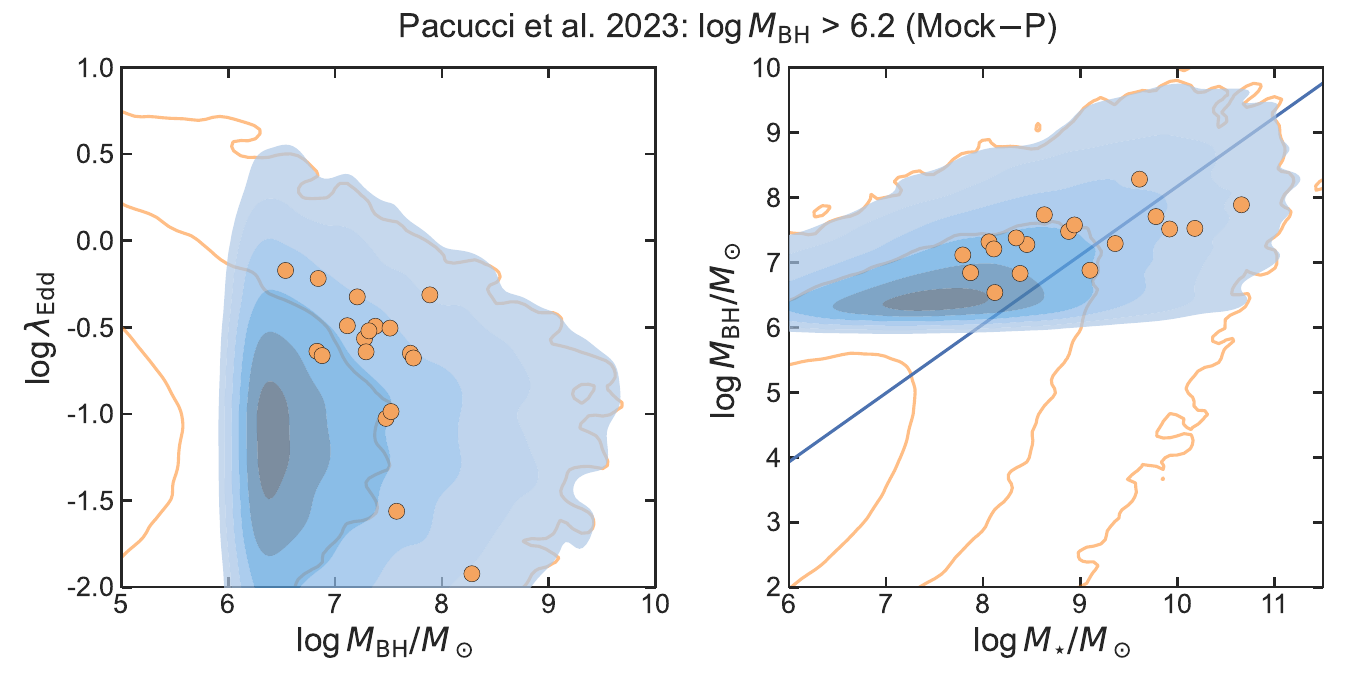}
\caption{Bivariate distribution of the Mock-L (top) and Mock-P (bottom) samples in the observed (i.e., including mass uncertainties) $\mbh-\edd$ and $\mbh-\m$ planes. Mock samples with and without incorporating selection functions ($\loglbol>44.1\ \ergs$ and $\rm FWHM>1000\ \kms$) are displayed as blue and orange contours, respectively. The observed JDEEP sample is represented by orange points. 
The distribution of $\mbh-\edd$ for mock AGNs with $\loglbol>44.7\ \ergs$ and $\rm FWHM>1100\ \kms$, matching the selection function of the \cite{Matthee2023} sample (pink points), is shown in the upper left inset.
The upper right inset presents the luminosity distribution of Mock-L AGNs with $7.0<\logm/\msun<7.3$ and $6.4<\logmbh/\msun<6.9$. Only BHs with $\loglbol>44.1\ \ergs$ and $\rm FWHM>1100\ \kms$ (blue shaded region, corresponding to $\logedd \gtrsim -0.46$) can be included in the final Mock-L sample. However, they all remain in the Mock-P sample since their BH masses exceed the detection limit ($\logmbh>6.2\ \msun$) in \cite{Pacucci2023}.
} 
\label{fig:bivariate}
\end{figure*}

\cite{Pacucci2023} conducted a statistical analysis of the same JDEEP sample as in this study and inferred an intrinsic mass relation of $\logmbh = 1.06_{-0.09}^{+0.09}\, \logm - 2.43_{-0.83}^{+0.83}$ with a scatter of $0.69_{-0.39}^{+0.69}$ dex. Their best-fit relation is plotted as a magenta dashed curve in Figure~\ref{fig:best_fit}, which is $\sim1.2$ dex above our default best-fit relation at $7.8<\logm/\msun<10.6$, and $\sim0.9-1.7$ dex above all the fitting results shown in Table \ref{table:params}. 
As a result, they concluded that the $\mbh/\m$ ratio at $z\sim5$ is intrinsically above that of local AGNs in \cite{Reines2015} (RV15) at a $>3\,\sigma$ confidence level, and the discrepancy is even more pronounced when comparing to the G20 relation (Figure \ref{fig:best_fit}). This led them to infer that the overall high-redshift BH population is overmassive by $\sim10-100$ times compared to their local counterparts (i.e., lack of undermassive BHs at high redshift). 

However, it is important to recognize the limitations of these local samples and acknowledge that their BH and stellar masses are not always consistently measured with the high-$z$ samples. As detailed in Appendix \ref{sec:appendix}, the \rel\ relation of local AGNs might be biased low (i.e., missing massive BHs, such as those hosted by massive quiescent galaxies in KH13) due to the potential mass-dependent \pac. The G20 sample, constructed by compiling local galaxies with mostly dynamical BH mass measurements, is highly inhomogeneous and incomplete, especially in the low-mass end. In fact, overmassive BHs ($\mbh/\m \sim 10\%$ at $\logm/\msun\sim9$) have also been suggested in the nearby universe \citep[e.g.,][]{Secrest2017, FerreMateu2021, Mezcua2023}, occupying a similar region in the \rel\ plane as high-redshift AGNs. However, such a parameter space is not covered by the G20 sample. It is thus essential to bear in mind that the local relation itself may be biased and affected by the complicated selection function when conducting evolutionary studies \citep{Schulze2011, Shankar2016}.

Aware of these caveats, we find that our best-fit relation aligns with both the RV15 and G20 relations within the $1\,\sigma$ uncertainty region. There is no evidence that the growth of the entire BH population outpaces their host galaxy at early cosmic times. The observed AGNs do systematically lie above the mean $\mbh-\m$ relation measured in the local universe due to selection effect, but they do not represent the whole population of AGNs at high-$z$. 

The discrepancy between our result and that of \cite{Pacucci2023} mainly arises from two differences. 
First, \cite{Pacucci2023} adopted a constant detection limit in BH mass (${\rm log}\,\mathcal{M}_{\rm BH, lim}/\msun\sim6.2$ at $z\sim5$), derived from the \ha\ flux limit ($\sim10^{-19}\,\rm erg\,s^{-1}\,cm^{-2}$) and a BL width of 1000\ \kms, via the average virial estimator. In contrast, our selection function is expressed in terms of AGN luminosity (${\rm log}\,L_{\rm bol, lim}/\ergs\sim44.1$) and BL width, while the detection of a given BH mass depends on their interplay with the underlying ERDF (Eq.~\ref{eq:selection}). Second, they allowed the model parameters to vary freely during fitting, whereas we only vary the model parameters within the permitted region. 

We use a simulation to demonstrate the distinctions between the two methodologies. We generate mock AGNs by randomly sampling Equation \ref{eq:p_mo_so} using the intrinsic \rel\ relations and sample selection functions from both works as done in Figure \ref{fig:sample_simu}, integrating measurement uncertainties for each observed source. 
The simulated samples (with parameters of $\mo$, $\so$, $w$, $l$, and $\lambda_{\rm o}$) based on the mass relation in this work (default result in Table \ref{table:params}) and  \cite{Pacucci2023} are denoted as Mock-L and Mock-P, respectively. 
To maintain consistency with the model assumptions in \cite{Pacucci2023}, we adopt the \cite{Song2016} SMF to simulate their mass relation. 

The resulting distribution of each mock sample in the observed $\mbh-\edd$ and $\mbh-\m$ planes with and without selection functions applied are represented by the orange and blue contours in Figure \ref{fig:bivariate}, respectively. 
Considering a small grid with $7.0<\logm/\msun<7.3$ and $6.4<\logmbh/\msun<6.9$, all the mock BHs in that bin are detectable in \cite{Pacucci2023} as their masses exceed ${\rm log}\,\mathcal{M}_{\rm BH, lim}/\msun\sim6.2$. However, only those with $\loglbol/\ergs>44.1$ and $\rm FWHM>1000\ \kms$ (see the right inset in Figure \ref{fig:bivariate}) can surpass our detection limit and be included in the Mock-L sample. Such a different treatment has dramatically altered the distribution of model BHs, thus affecting the evaluation of the likelihood of the observed sample; even though the average detection limit on \mbh\ is somewhat similar in both works.

Notably, the Mock-L sample (blue contours) successfully replicates the distribution of observed AGNs in both the $\mbh-\edd$ and $\mbh-\m$ planes, demonstrating that only BHs above a limit of $\lbol \propto \mbh \times \edd$ can be detected. Our results align well with the predictions from the CAT semi-analytic model \citep{Trinca2023}, which suggests that AGNs detected in a CEERS-like survey ($\loglbol/\ergs>44.4$) peak within the mass range of $6<\logmbh/\msun<8$ and $8<\logm/\msun<10$.
The slight deviation in the $\mbh-\edd$ plane occurs because these samples are not strictly flux-limit selected; instead, they involve complex target priority assignment during shutter allocation. Consequently, the luminosity distribution of mock AGNs does not precisely match the observed samples, with the latter having a slightly higher mean. In the top left panel of Figure \ref{fig:bivariate} we also display the $\mbh-\edd$ distribution of mock AGNs selected with $\loglbol/\ergs>44.7$ and $\rm FWHM>1100\ \kms$, matching the flux-limited BL AGN sample in \cite{Matthee2023} selected from slitless spectroscopy. It can be seen that with a shallower luminosity limit, the $\mbh-\edd$ distribution is shifted towards higher BH mass. Intriguingly, our model well reproduces the distribution of the \cite{Matthee2023} sample which has a much more uniform sample selection criteria, thereby validating our assumed ERDF and fitting result. 

Conversely, the Mock-P sample vastly overpredicts the number of ``detectable'' low-\mbh, low-\edd\ AGNs, which are actually not observed in the real samples. Additionally, although the Mock-P sample generally follows the average \rel\ relation of the observed data points as it is optimized for in the fitting, its highest density region (peaks at $7<\logm/\msun<8$, see also Section 4.3 in \citealt{Pacucci2023}) deviates significantly from the observed sample. This is because the statistical model in \cite{Pacucci2023} ignores the substantial scatter associated with the virial estimator when deriving the BH mass limit from it (see Eq.~\ref{eq:sigmaw}). Additionally, it does not consider the fact that a low-mass BH with a high \edd\ can still surpass the flux limit, while a massive BH with a low \edd\ can be undetectable.
This population of low-\mbh, low-\edd\ AGNs, hosted by $\logm/\msun<8$ galaxies, is responsible for the huge number of ``detectable'' but missing AGNs with ${\rm log}\,\mathcal{M}_{\rm BH, lim}<\logmbh<6.5$ reported in \cite{Pacucci2023}. Our simulation and comparison result demonstrate that adopting a constant detection limit in \mbh\ is an inadequate representation of the sample selection function. It is imperative to self-consistently integrate the ERDF in the statistical model to accurately assess the impact of the flux limit on the detection limit of BH mass. 

\subsection{Implications for JWST studies of BH seedings}
\label{subsec:seeding}

Our modeling of selection biases underscores that the observed AGNs only represent the tip of the iceberg of the underlying BH population. The inferred intrinsic mass relation in the low-mass regime suggests the potential existence (albeit with large uncertainties) of a substantial population of low-mass BHs with $\logmbh/\msun\lesssim5$ that remain undetected or unidentified in the current JWST surveys. However, we acknowledge that their presence is inferred by assuming a linear \rel\ relation with a mass-independent Gaussian scatter, which may not be the case \cite[e.g.,][]{Habouzit2022}. 

\begin{figure}
\centering
\includegraphics[width=\linewidth]{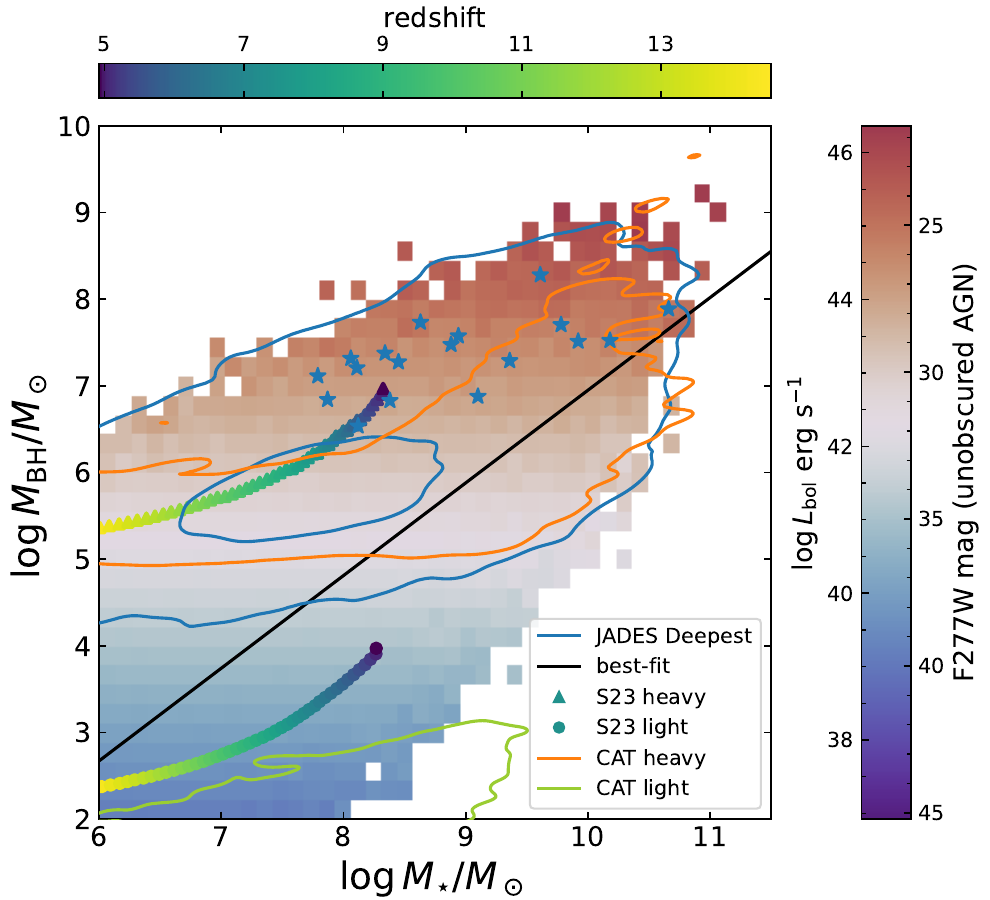}
\caption{The comparison of the intrinsic \rel\ relation with theoretical predictions at $z\sim5$. The blue stars represent the observed JDEEP AGNs. The squares, color-coded by the average \lbol\ and the corresponding F277W magnitude in each mass bin (true masses without adding uncertainties), represent mock AGNs simulated using Equation \ref{eq:p_mo_so} based on our default  intrinsic mass relation (black solid line with a scatter of 0.91 dex).  The green triangles (top) and circles (bottom) show the evolutionary tracks (color-coded by redshift) of heavy ($10^5\,\msun$) and light ($100\,\msun$) seeds based on the $\tau_{\rm fold}=225\,\rm Myr$ model in \cite{Scoggins2023seed}, corresponding to BHs accreting at the mean \edd\ of JDEEP AGNs. The \rel\ relations of the descendants of heavy and light seeds evolved to $z\sim5$ in the Eddington-limited CAT model are depicted as orange and green contours ($3\,\sigma$ level), respectively. The predicted distribution of BHs detectable in the JADES Deepest survey ($\rm F277W < 30.5$~mag and $\loglbol \gtrsim 42.6\ \ergs$) under idealized scenarios (i.e., assuming no dust extinction and host galaxy contamination, with stellar and BH masses being accurately measured, even at $\rm FWHM < 1000\,\kms$) is shown as blue contours ($1\,\sigma$ and $3\,\sigma$ levels).} 
\label{fig:seeding}
\end{figure}

Missing a significant population of low-mass BHs has important consequences for our understanding of BH seeding mechanisms and the earliest coevolution of SMBHs and their host galaxies. 
In Figure \ref{fig:seeding}, we display the assembly history of heavy seed ($10^5\,\msun$) and light seed ($100\,\msun$) seeded at $z=25$ and grown to a final stellar mass of $\logm/\msun\sim 8.3$ at $z\sim5$ with an $e$-folding time of $\tau_{\rm fold}=225\,$Myr based on the semi-analytic model described in \cite{Scoggins2023seed} (S23). The $\tau_{\rm fold}$ is chosen such that, on average, BHs are accreting at the mean Eddington ratio (0.2) of JDEEP AGNs, meaning that we are approximating the population-averaged accretion rate to the time-averaged one. The stellar mass is instead derived from scaling relations with the halo mass following \cite{Wise2014} and \cite{Behroozi2019}, and no self-regulated feedback between BHs and galaxies is included. The comparison between the evolutionary track of model seed BHs with our intrinsic \rel\ relation demonstrate that, while the observed overmassive AGNs may originate from heavy seeds that ultimately reach $\logmbh/\msun \sim 7$ at $z\sim5$ \citep[e.g.,][]{Pacucci2023, Narayanan2023, Bogdan2024}, the lighter seeds are responsible for the formation of the bulk of low-mass BHs ($\logm/\msun \lesssim 5$) below the average mass relation. 

Theoretically, the existence of this low-mass population depends on the effective growth of light and medium-weight seeds. \cite{Smith2018} explored the evolutionary track of light seeds within the framework of the {\it{Renaissance}} simulation and found that the growth of light seeds is impeded by their random motion and distribution in the host galaxy. Furthermore, although dense gas clumps are abundant in many mini-haloes, they are either consumed by star formation and destroyed by stellar feedback, or the light seeds are too small to settle down into such clumps (see also \citealt{Alvarez2009}). 
A similar conclusion has been reached in the Eddington-limited CAT model \citep{Trinca2022, Schneider2023}, where the inefficient growth of light seeds results in a distinct mass gap between the descendants of light and heavy seeds in the \rel\ plane, as shown in Figure~\ref{fig:seeding}. 

However, in models that allow episodic super-Eddington accretion \citep{Trinca2022, Schneider2023}, or if ubiquitous cold dense gas flows in the early universe can facilitate continuous but chaotic accretion (thus reducing the BH spin and radiative efficiency) at a sufficiently high rate \citep{Alexander2014, Zubovas2021}, 
or if early BH growth is driven by their frequent mergers \citep{Bhowmick2024},
even light seeds can rapidly accumulate their BH masses. 
In the case of medium-weight seeds forming in stellar clusters, their growth is expected to be easier compared to light seeds, because they can consume gas in the stellar clusters at super-Eddington rates \citep{Alexander2014} and their dynamics is stabilized by the dense stellar envelope, facilitating both their growth and their mergers \citep{Biernacki2017}.
These efficient growth mechanisms result in a continuous and elevated mass relation that matches the observed overmassive AGNs \citep{Trinca2022, Bhowmick2024}.

Interestingly, the average \rel\ relation evolved from heavy seeds in the Eddington-limited CAT model, representing BHs growing from ``typical'' evolutionary pathways (i.e., no extreme conditions are required to form significant outliers in \mbh/\m), aligns well with our best-fit relation within the observed stellar mass range. However, more efficient and diverse growth for a portion of seed BHs, possibly resulting from a varying seeding environment, formation mechanism, and assembly history (e.g., a combination of Eddington-limited and super-Eddington accretion onto different seeds, as well as mergers), is required to explain our mass relation with a much larger scatter towards both massive and low-mass BHs at a given stellar mass. Observations directly probing the $\logmbh/\msun<5$ regime are crucial for validating our intrinsic \rel\ relation and distinguishing between theoretical models.

In Figure \ref{fig:seeding}, we present the predicted observable \rel\ relation for the JADES Deepest survey \citep{Eisenstein2023} under {\it{idealized}} conditions at $z\sim5$ as blue contours ($1\,\sigma$ and $3\,\sigma$ levels). This is achieved by applying the selection function, which represents the detection limit in AGN bolometric luminosity converted from the $10\,\sigma$ depth in the F277W band ($\rm F277W \lesssim30.5$~mag), to our intrinsic \rel\ relation. In this experiment, all AGNs above the detection limit are assumed to be free from dust extinction and can be effectively identified, with their BH and stellar masses accurately measured. We acknowledge that this prediction is based on the assumption that low-mass BHs follow the same ERDF as their more massive counterparts, which may not hold true.  While the Deepest JADES survey has the capability to probe luminosities down to $\loglbol/\ergs \sim 42.6$ and unveil BHs with $\logmbh/\msun \sim4$ accreting at the Eddington limit, the rarity of such objects (lie outside the $3\,\sigma$ blue contour in Figure \ref{fig:seeding}) implies that the majority of detectable BHs will likely have $5<\logmbh/\msun<6$ and $6<\logm/\msun<8$, accreting at $\logedd \sim -0.7$. Consequently, the identification of light and medium-weight seeds remains an extremely challenging task even with the deepest JWST surveys. Their existence might only be revealed by detecting gravitational wave emissions produced by seed BH mergers \cite[e.g.,][]{Sesana2009, Ricarte2018, Bhowmick2024}.

It is worth noting that detecting a flat BH distribution at $5<\logmbh/\msun <6$ and $6<\logm/\msun <8$ in JADES does not necessarily verify the prediction from the Eddington-limited CAT model (i.e., a gap and flattening at $\logmbh\sim5$ in Figure \ref{fig:seeding}). This is because due to the finite detection limit, the observable BHs predicted by our intrinsic mass relation, which is very different from the CAT model at $\logm/\msun <8$, exhibits a nearly identical flat distribution in the \rel\ plane (see the $1\,\sigma$ blue contour in Figure \ref{fig:seeding}).

In summary, caution is warranted when inferring the dominant formation channel of high-redshift SMBHs from biased samples and comparing observations with theoretical models, as ``undermassive'' AGNs originating from lighter seeds  remain largely undetectable \citep{Volonteri2023}.

\section{Conclusions}
\label{sec:conclusion}
In this paper, we investigated the impact of selection biases (i.e., finite detection limit and requirements on detecting BLs) and measurement uncertainties on the observed \rel\ relation at high redshift. We demonstrated that these factors can fully account for the observed flattening and enhancement in the ratio of $\mbh/\m$ compared to local values for both JWST-discovered low-luminosity AGNs and the most luminous quasars at $4<z<7$ (Section \ref{subsec:observed_offset}). 

Moreover, by forward modeling the observed low-luminosity AGN sample, we inferred an intrinsic, bias-corrected mass relation in the low-mass regime, parameterized as $\logmbh = 1.07_{-0.40}^{+0.55}\, \logm - 3.75_{-5.35}^{+5.08}$ with a scatter of $0.91_{-0.40}^{+0.55}$. In contrast to \cite{Pacucci2023}, who concluded that the overall high redshift BH population are intrinsically overmassive by $\sim10-100$ times than their local counterparts (i.e., lack of undermassive BHs at high-$z$), our careful treatment of the selection bias results in an intrinsic \mbh/\m\ $\sim1.2$ dex below the observed, biased sample and is consistent with local values (Sections \ref{subsec:intrinsic} and \ref{subsec:compare}). Our results underscore that the observed AGNs only populate the upper envelope of the underlying BH population, even in deep JWST fields, and demonstrate the importance of properly considering the scatter of the mass relation when investigating its potential evolution.

The inferred intrinsic mass relation demonstrates that a significant population of low-mass BHs ($\logmbh/\msun < 5$), potentially formed from lighter seeds, might be missing even in the deepest JWST surveys (Section \ref{subsec:seeding}). Meanwhile, the elevated scatter for the high-redshift \rel\ relation in the low-mass regime, in comparison to the KH13 relation and massive quasars at $z\lesssim2$ \citep{Li2021mass, Tanaka2024}, might be indicative of a merger-averaging origin for establishing the close connections between massive BHs and bulges in the local universe \citep{Peng2007, Jahnke2011}. 

Nevertheless, inferring the intrinsic mass relation from the observed, and likely strongly biased sample is a challenging task fraught with substantial uncertainties.
Extending the current ``effective'' luminosity limit to the theoretical minimum achievable in existing JWST surveys is essential for obtaining more precise constraints on the low-mass population. The secondary BHs of the three dual AGN candidates in \cite{Maiolino2023}, whose average \mbh\ is $\sim1.3$ ~dex smaller than the BL AGN sample studied here, support the potential existence of an abundant population of lower-mass BHs that are reachable with JWST.
Significant progress has also been made in identifying type 2 AGN candidates at $z>4$ with luminosities $\sim10$ times lower than the BL AGN sample, primarily through the detection of high-ionization lines \citep{Scholtz2023}. These advancements need to be assisted with the establishment of alternative methods (e.g., via narrow line ratios or variability; \citealt{Baron2019, Sun2023}) for estimating BH masses in low-mass AGNs, where the narrowed or extincted ``broad" lines are no longer distinguishable from those powered by star formation processes.
This, in turn, will provide valuable insights into constraining the undetectable population through statistical modeling, as done in this work and others \citep[e.g.,][]{Schulze2014, Li2021mass, Wu2022}. 

However, this bias-correction approach inevitably depends on the assumed forms of the intrinsic mass relation and the underlying distribution functions. Obtaining independent measurements of the ERDF, AGN fraction, and duty cycle (e.g., via clustering measurements; \citealt{Shen2007}) in the low-mass regime and improving theoretical predictions on the assembly history of BH seeds would be crucial for refining our model assumptions, narrowing down the fitting parameter space, and reducing the posterior uncertainties.

On the other hand, it can be seen that the JDEEP and SHELLQs surveys have started to probe BHs positioned on the local relation and the intrinsic mass relation inferred in this study at the massive stellar mass end. However, such massive galaxies are rare in pencil beam surveys, thus the current number counts is insufficient for statistical analysis. Expanding deep JWST surveys to cover a significantly larger area, or extending JWST observations of SHELLQs quasars to even fainter systems would provide the opportunity to directly probe undermassive BHs in massive galaxies at early cosmic times. This would allow for a more robust constraint on the intrinsic \rel\ relation at the massive end, mitigating the impact of selection bias to a greater extent.

\begin{acknowledgements}
We thank Raffaella Schneider and Alessandro Trinca for providing us the model predictions from CAT. 
J.S. is supported by JSPS KAKENHI (JP22H01262) and the World Premier International Research Center Initiative (WPI), MEXT, Japan. This work was supported by JSPS Core-to-Core Program (grant number: JPJSCCA20210003). This work is partially supported by JWST programs JWST-GO-02057 and JWST-AR-03038 through a grant from the STScI under NASA contract NAS5-03127.
\end{acknowledgements}

\appendix
\section{Impact of a BH-mass dependent BL AGN fraction}
\label{sec:appendix}

\cite{Schulze2011} first illustrated that a mass-dependent \pac\ induces a further bias to the observed mass relation. This arises from the fact that high redshift observations can only detect BHs and measuring their masses when they are activated as BL AGNs. The activation efficiency and duration, and hence \pac, depends on the complex interplay between the self-regulated fueling and feedback of BHs with their host galaxy and the gas reservoir, and is likely to evolve strongly with redshift as well as BH and stellar masses. Consequently, the mass relation traced by BL AGNs is {\it{not}} equivalent to that of the entire BH population. A constant \rel\ relation coupled with a redshift- and mass-dependent active fraction could result in an evolving mass relation for AGNs \citep{Volonteri2011, Schulze2011}. 

In this Appendix, we investigate the impact of a BH mass-dependent \pac\ on the observed mass relation. Given the lack of observational constraint at high redshift, we simply express the mass-dependency as $\pac\,(m) \propto (0.1\, m - 0.4)^\gamma$ (arbitrary normalization, which will be canceled out in Eq.~\ref{eq:mo_so}; negative values are set to 0), where $\gamma$ can take values of 1 and 5 (Figure \ref{fig:pac}a). This somewhat arbitrary form is chosen to mimic the increased fueling efficiency and duty cycle of massive BHs at high-redshift \citep{Aversa2015, Zhu2022}, which is required to explain their large BH masses assembled within the first billion year of the universe. Following the procedure outlined in Section \ref{subsec:impact} and Figure \ref{fig:sample_simu}, we present in Figure \ref{fig:pac}b the predicted observed mass relation for the entire $z\sim5$ BL AGN population (i.e., including those below the luminosity limit) by incorporating $\Omega(m) = \pac\,(m)$ into Equation~\ref{eq:p_mo_so}, assuming that the underlying BH population follows the KH13 relation. Both a constant and a BH mass-dependent \pac\ are displayed for comparison. 

\begin{figure*}
\centering
\includegraphics[width=\linewidth]{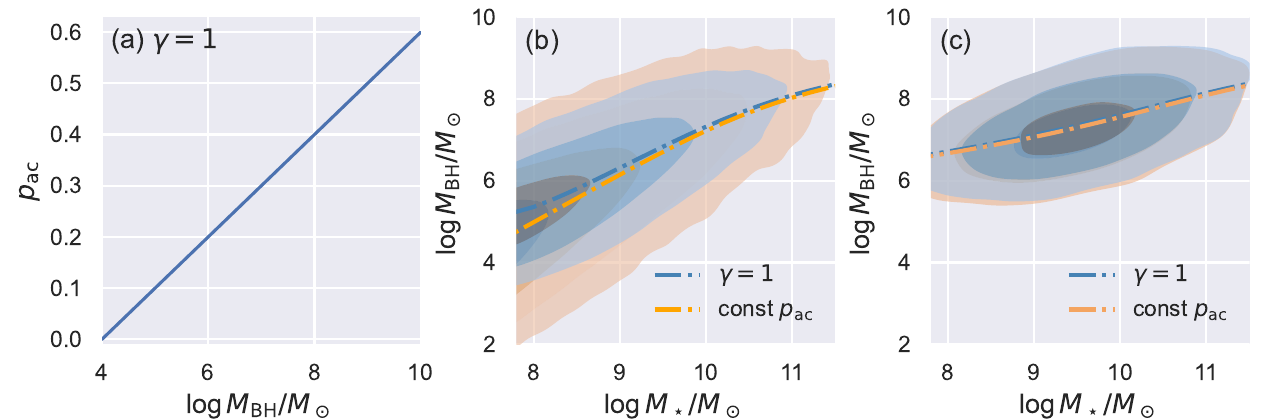}
\includegraphics[width=\linewidth]{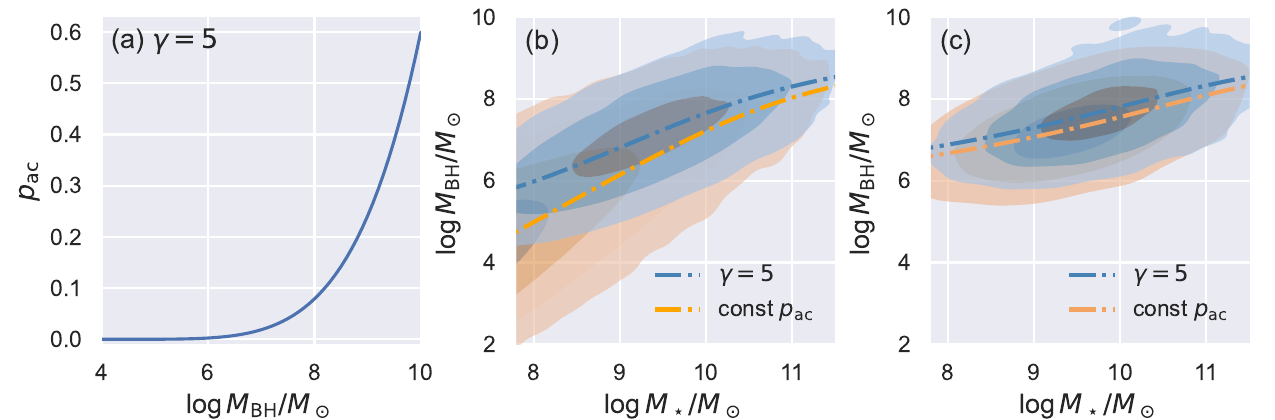}
\caption{Impact of a BH mass-dependent BL AGN fraction (\pac) on the observed (i.e., incorporating mass uncertainties with $\smo = 0.35$ and $\sso = 0.45$) mass relation at $z\sim5$. {\bf{Panel (a)}}: \pac\ as a function of BH mass in the form of $\pac \propto (0.1\,m - 0.4)^\gamma$ with $\gamma=1$ and $\gamma=5$. {\bf{Panel (b)}}: The predicted observed mass relation for the entire BL AGN population (i.e., including those below the flux limits) at $z\sim5$ based on a constant (orange) and a mass-dependent \pac\ (blue), assuming that the underlying BH population follows the KH13 relation. The bivariate \rel\ distribution is shown as contours, while the mean relation is represented by dashed curves. {\bf{Panel (c)}}: Similar to Panel (b) but for detectable BL AGNs with $\loglbol>44.1\ \ergs$ and $\rm FWHM>1000\ \kms$. 
}
\label{fig:pac}
\end{figure*}

It can be seen that the predicted mass relation for BL AGNs is biased upwards when a mass-dependent \pac\ is adopted, even without invoking any selection on AGN luminosity. This is caused by assigning more weight to massive BHs through \pac. The effect is particularly pronounced for the $\gamma=5$ model, where \pac\ increases dramatically with increasing $\mbh$. 

We then incorporate the selection function of the JDEEP sample to predict the observed mass relation for detectable AGNs (Figure \ref{fig:pac}c). Interestingly, the resulting mass relation of a constant and a mass-dependent \pac\ model becomes similar. The median difference of the two mass relations is 0.05~dex for $\gamma=1$ and 0.25~dex for $\gamma=5$. This is because the \mbh\ distribution for BHs surpassing the selection criteria is limited to a relatively narrow range (spanning $\sim1$ dex for the highest density region), where \pac\ does not vary strongly with the BH mass.

\begin{figure*}
\centering
\includegraphics[width=\linewidth]{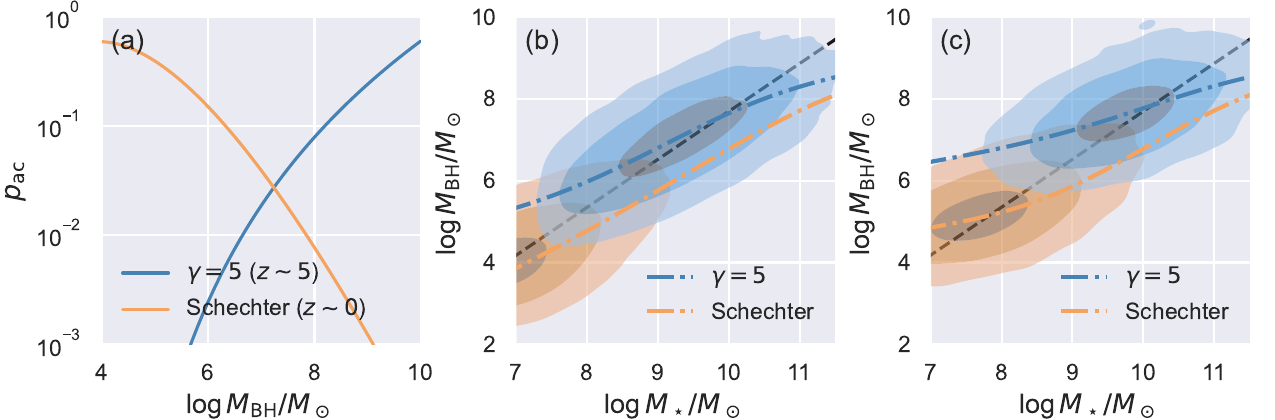}
\caption{The distinct impact of a redshift- and BH-mass dependent BL AGN fraction (Panel a) on the observed mass relation at $z\sim5$ and $z\sim0$. The underlying BH population at both redshifts are assumed to follow the KH13 relation (black dashed line). The blue curves and contours mirror the $\gamma=5$ model in Figure \ref{fig:pac}, showcasing the predicted mass relation for BL AGNs at $z\sim5$ after incorporating the mass-dependent \pac\ (Panel b) together with the luminosity and line width cut (Panel c) to the KH13 relation. The orange curves and contours depict the predicted mass relation for $z\sim0$ BL AGNs by replacing the $\gamma=5$ \pac\ model to the Schechter model in \cite{Ananna2022}, and adjusting the selection function to $\loglbol > 42.0\ \ergs$ and $\rm FWHM>500\ \kms$ for local AGNs.}
\label{fig:pac_local}
\end{figure*}

The situation might be reversed in the local universe, as the accretion onto the most massive BHs has almost ceased. In Figure \ref{fig:pac_local} we investigate the impact of a redshift- and BH-mass dependent \pac\ on the observed mass relation for BL AGNs at $z\sim5$ and $z\sim0$, again under the assumption that the underlying BH population intrinsically follows the KH13 relation. We adopt the $\gamma=5$ model for $z\sim5$ AGNs. The $\pac(m)$ in the local universe is adopted from \cite{Ananna2022} based on Swift/BAT AGNs, which can be approximated by a Schechter function that decreases strongly at the massive end (Figure \ref{fig:pac_local}a). As shown in Figure \ref{fig:pac_local}b, the predicted mass relation for local BL AGNs is biased towards low-mass BHs based on the Schechter \pac. Its discrepancy with high-redshift AGNs becomes even prominent when the selection function, represented by $\loglbol/\ergs>44.1$ and $\rm FWHM>1000\ \kms$ at $z\sim5$, and $\loglbol/\ergs>42.0$ and $\rm FWHM>500\ \kms$ at $z\sim0$ \citep{Reines2015}, is incorporated (Figure \ref{fig:pac_local}c). Therefore, at least part of the offset between high-$z$ and local AGNs seen in \cite{Pacucci2023} and \cite{Stone2023} can be attributed to the mass-dependent \pac.   

We emphasize that the exact form of $\pac(m)$ is still highly uncertain, and thus our results based on two extreme cases of \pac\ at $z\sim0$ and $z\sim5$ should be treated as illustrative only. Nevertheless, our analysis underscores that BL AGNs are biased tracers of BHs, which must be kept in mind when adopting AGNs as the local baseline to study the evolution of the mass relation. 


\bibliography{sample631.bbl}{}
\bibliographystyle{aasjournal}

\end{document}